\documentclass{article}

\usepackage{PRIMEarxiv}

\usepackage[utf8]{inputenc} % allow utf-8 input
\usepackage[T1]{fontenc}    % use 8-bit T1 fonts
\usepackage{hyperref}       % hyperlinks
\usepackage{url}            % simple URL typesetting
\usepackage{booktabs}       % professional-quality tables
\usepackage{amsfonts}       % blackboard math symbols
\usepackage{nicefrac}       % compact symbols for 1/2, etc.
\usepackage{microtype}      % microtypography
\usepackage{lipsum}
\usepackage{fancyhdr}       % header
\usepackage{graphicx}       % graphics
\graphicspath{{media/}}     % organize your images and other figures under media/ folder
\usepackage{amsmath}
\usepackage{natbib}
\usepackage[separate-uncertainty=true,multi-part-units=single]{siunitx}  
\usepackage{xcolor}
\usepackage{caption}
\usepackage{subfig}
\usepackage[version=4]{mhchem}
\usepackage{xspace}
\usepackage{bm}

\DeclareSIUnit\bar{bar}
\DeclareSIUnit\debye{D}
\DeclareSIUnit\angstrom{\textup{Å}}

\newcommand{\SPTPCSAFT}{SPT$_\mathrm{PC\mh SAFT}$\xspace}
\newcommand{\feos}{FeO$_\mathrm{s}$\xspace}
\DeclareMathSymbol{\mh}{\mathord}{operators}{`\-}

\title{Understanding the language of molecules:\\ Predicting pure component parameters for the PC-SAFT equation of state from SMILES
}

\author{
    Benedikt Winter\\
    ETH Zurich, EPSE \\
    Email: \href{mailto:bewinter@ethz.ch}{bewinter@ethz.ch}
    \and
    \textbf{Philipp Rehner}\\
    ETH Zurich, EPSE \\
    Email: \href{mailto:prehner@ethz.ch}{prehner@ethz.ch}
    \and
    \textbf{Timm Esper}\\
    University Stuttgart, ITT \\
    Email: \href{mailto:esper@itt.uni-stuttgart.de}{esper@itt.uni-stuttgart.de}
    \and 
     \\
    \textbf{Johannes Schilling}\\
    ETH Zurich, EPSE \\
    Email: \href{mailto:jschilling@ethz.ch}{jschilling@ethz.ch}
    \and 
    \\
    \textbf{Andr\'e Bardow}\thanks{Corresponding author. Email: \href{mailto:abardow@ethz.ch}{abardow@ethz.ch}}\\
    ETH Zurich, EPSE \\
    Email: \href{mailto:abardow@ethz.ch}{abardow@ethz.ch}
}

\begin{document}
\maketitle

\begin{abstract}

A major bottleneck in developing sustainable processes and materials is a lack of property data. Recently, machine learning approaches have vastly improved previous methods for predicting molecular properties. However, these machine learning models are often not able to handle thermodynamic constraints adequately. In this work, we present a machine learning model based on natural language processing to predict pure-component parameters for the perturbed-chain statistical associating fluid theory (PC-SAFT) equation of state. The model is based on our previously proposed SMILES-to-Properties-Transformer (SPT). By incorporating PC-SAFT into the neural network architecture, the machine learning model is trained directly on experimental vapor pressure and liquid density data. Combining established physical modeling approaches with state-of-the-art machine learning methods enables high-accuracy predictions across a wide range of pressures and temperatures, while maintaining the physical meaning of PC-SAFT parameters. \SPTPCSAFT demonstrates exceptional prediction accuracy even for complex molecules with various functional groups, outperforming traditional group contribution methods by a factor of four in the mean average percentage deviation. Moreover, \SPTPCSAFT captures the behavior of stereoisomers without any special consideration. To facilitate the application of our model, we provide predicted PC-SAFT parameters of more than \num{13645} components, making PC-SAFT accessible to all researchers.
\end{abstract}

% keywords can be removed
 \keywords{PC-SAFT \and machine learning \and computational chemistry}

\section{Introduction}

Developing advanced materials like chemical products, fuels, or refrigerants is vital for sustainable solutions in various industries. To achieve this goal, designing new molecules with tailored properties is crucial. However, exploring all possible molecules experimentally is impossible, given the vast array of potential molecular candidates. As a result, models are needed that can rapidly predict molecular properties to streamline the molecular discovery and development of sustainable products and processes.

Over the years, the research on predicting molecular properties has led to many approaches based on, e.g., quantitative structure-property relationships (QSPRs) \citep{Katritzky.1995,Hughes.2008}, group contribution (GC) methods \citep{Fredenslund.1975,Marrero.2001,Hukkerikar.2012, Sauer.2014} and quantum mechanics \citep{Klamt.1995,Lin.2002,Schleder.2019}. However, many of these classical methods either have low accuracy, are limited to certain functional groups, or require large computational resources. As a recent addition to these approaches, machine learning methods have emerged as a powerful tool due to their ability to learn complex patterns and generalize from data, overcoming some of the shortcomings of the classical methods. Some recent examples of machine learning approaches include methods for the prediction of binary properties such as activity coefficients \citep{Jirasek.2021,Winter.2022,SanchezMedina.2022,Rittig.2023} or a large range of pure component properties \citep{Liu.2019,Venkatasubramanian.2019,Ding.2021,Alshehri.2021}. 

However, the majority of recent machine learning approaches focus on singular properties, not a holistic description of a system.  Thermodynamics teaches that equilibrium properties of fluids are not independent but rather related through an equation of state. Modern equations of state are expressed as a thermodynamic potential, usually the Helmholtz energy, as a function of its characteristic variables. All equilibrium properties are then available as partial derivatives of the thermodynamic potential. Equations of state can be broadly classified into three categories: 1) cubic equations of state (such as the Peng-Robinson \citep{Peng.1976} and the Soave-Redlich-Kwong \citep{Soave.1972} equation of state), 2) highly accurate reference equations for specific systems (including water \citep{Wagner.2002}, carbon dioxide \citep{Span.1996}, nitrogen \citep{Span.2000}, and natural gas components \citep{Kunz.2012}), and 3) molecular equations of state (such as the SAFT family \citep{Chapman.1990,Gross.2001,Llovell.2004,Lafitte.2013}). The main distinction among these categories lies in the data required for parameterization, with cubic equations of state necessitating the fewest parameters and reference equations of state demanding the most.

Parameterizing equations of state typically relies on experimental data, which is often unavailable for novel molecules or expensive to obtain from commercial databases or experiments. In the absence of experimental data, various predictive methods have been developed for equations of state, primarily focused on GC methods \citep{Shaahmadi.2023,Privat.2023}. Since group contribution methods rely on a predefined set of functional groups and their respective contributions, those methods are limited to certain subsets of the molecular space and often struggle to predict the properties of more complex molecules accurately. Furthermore, capturing effects linked to isomers or more intricate intermolecular forces requires the definition of higher-order groups, for which adequate parametrization is more data-demanding\citep{Gani.2019}.

Recently, machine learning (ML) methods have been developed to predict pure component parameters for equations of state. The focus has been on the perturbed-chain statistical associating fluid theory (PC-SAFT) equation of state developed by \citet{Gross.2001}. The ML models use as input either group counts \citep{Matsukawa.2021}, molecular fingerprints \citep{Habicht.2023}, or a variety of molecular descriptors \citep{Felton.2023}. However, these methods are not trained directly on experimental property data but on previously fitted pure component parameters of PC-SAFT. This reliance on previously fitted pure component parameters vastly constraints the amount of available training data, thus likely limiting the applicability domain of these models. Moreover, small errors in predicted pure component parameters can have large effects on the final predicted fluid properties. Consequently, training machine learning models directly on experimental property data is preferred. 

In previous work, we demonstrated how explicit physical equations could be integrated into a machine learning framework, using the NRTL-equation as an example \citep{winter.2023}. However, integrating PC-SAFT into a machine learning framework presents two additional challenges: Firstly, PC-SAFT is not explicit in measurable properties like vapor pressures and liquid densities. Instead, vapor pressures and liquid densities have to be determined iteratively from partial derivatives of the Helmholtz energy, requiring a more sophisticated approach than a straightforward integration into the neural network. Secondly, the physical significance of the pure component parameters of PC-SAFT is the basis of its robust extrapolation, in particular to mixtures. Therefore, any predictive method should ensure that parameters meaningful for their physical basis are obtained.    

In this work, we present a natural language-based machine learning model for predicting pure component parameters of PC-SAFT trained directly on experimental data. For this purpose, the PC-SAFT equation of state is directly integrated into our previously proposed SMILES-to-Properties-Transformer (SPT) \citep{Winter.2022, winter.2023}. The resulting SPT-PC-SAFT model exhibits high prediction performance, accurately predicting thermophysical properties for complex molecules with various functional groups. Remarkably, our model is also capable of correctly predicting the behavior of stereoisomers.

\section{The SPT-PC-SAFT model}
\label{sec:methods}

The \SPTPCSAFT model is designed to allow the inclusion of explicit systems of equations into machine learning frameworks to apply physical constraints. This work uses the PC-SAFT equation of state, though other equations of state or any other system of equations could be integrated. SPT is a natural language processing model that utilizes the SMILES code of a molecule as input. Conceptually, our SPT model can be interpreted as an advanced group contribution approach that uses characters in the SMILES code as atomic groups and dynamically assembles higher-order groups via natural language processing.

\begin{figure}[htbp]
    \centering
    \includegraphics[width=\textwidth]{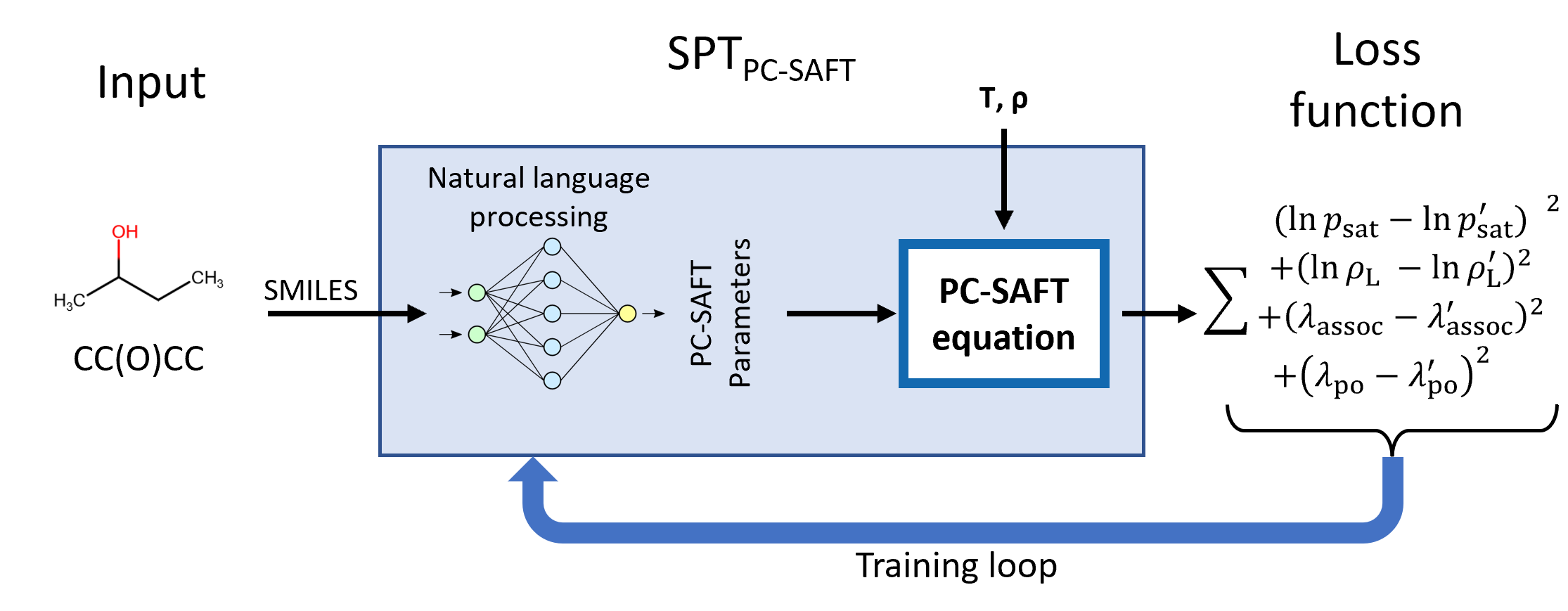}
    \caption{Overarching structure of the \SPTPCSAFT model and training. Molecules are represented as SMILES and passed into a natural language model to predict PC-SAFT parameters, which are, in turn, used to calculate vapor pressures $p_\mathrm{sat}$ and liquid densities $\rho_\mathrm{L}$ for a given temperature or temperature and pressure, respectively. Furthermore, the likelihood of molecules having associating ($\lambda_\mathrm{assoc}$) or polar ($\lambda_\mathrm{polar}$) interactions is predicted. During training, the loss function, i.e., target function, is calculated based on the natural logarithm of the pressure or density and the association and polarity likelihoods.}
    \label{fig:overall}
\end{figure}

Figure \ref{fig:overall} illustrates the overall structure of the proposed \SPTPCSAFT model: First, molecules are represented as SMILES codes, which are fed into a natural language processing model that predicts parameters, which are used within the PC-SAFT equation of state to compute vapor pressures and liquid densities at a given temperature or temperature and pressure, respectively. To preserve the physical meaning of the PC-SAFT parameters, the likelihood that a component is associating ($\lambda_\mathrm{assoc}$) or polar ($\lambda_\mathrm{polar}$) is also predicted by \SPTPCSAFT and molecules are only assigned associating or polar parameters if the molecule is predicted to be associating or polar. During the model training, the PC-SAFT equation of state is incorporated into the backward pass, allowing for the calculation of analytical gradients of the loss (target function) with respect to the model parameters. This integration enables us to train a machine learning model end-to-end on experimental data and not only on previously fitted parameters. 

In the following sections, the model and training procedure of \SPTPCSAFT are described in detail: Section~\ref{sec:architecture} introduces the architecture of the machine learning model and the integration of the PC-SAFT equation. Section \ref{sec:data} describes the data sources, data processing, and the definition of training and validation sets. In Section \ref{sec:hyper}, we describe the selection of hyper-parameters and the training process of \SPTPCSAFT.

\subsection{Model architecture}
\label{sec:architecture}

The model architecture of \SPTPCSAFT (Figure~\ref{fig:network}) is largely based on our previous SPT models \citep{Winter.2022, winter.2023}, which are in turn based on the natural language model GPT-3 \citep{NEURIPS2020_1457c0d6} using a decoder-only transformer architecture implemented by \citet{NIPS2017_3f5ee243}. The transformer architecture has been shown suitable for understanding not only the grammar of natural language but also the molecular grammar embedded within SMILES codes, a linear text-based molecular representation introduced by \citet{Weininger.1988}, leading to many successful applications in the field of chemistry \citep{Schwaller.2019,Honda.12.11.2019,Lim.2021,Kim.2021}. 

In the following, we present the \SPTPCSAFT architecture in three sections: input embedding (Section~\ref{sec:input_embedding}), multi-head attention (Section~\ref{sec:MHA}), and head (Section~\ref{sec:head}).

\subsubsection{Input embedding}\label{sec:input_embedding}

\SPTPCSAFT predicts thermodynamic equilibrium properties as calculated from PC-SAFT and the corresponding pure component parameters using the SMILES codes of a molecule as input. The SMILES code \citep{Weininger.1988} has become a widely adopted molecular representation for machine learning applications in chemical engineering and has been used in numerous recent studies \citep{Honda.12.11.2019,Wang.2019,Schwaller.2019,Lim.2021}. The SMILES code offers a linear string representation for complex branched and cyclic molecules. In the SMILES codes, atoms are denoted by their periodic table symbols, such as the character "N" for nitrogen, while hydrogen atoms are implicitly assumed. While single bonds are also implicitly assumed, double and triple bonds are indicated by the characters "=" and "\#", respectively. Branches are enclosed in brackets, and connections of ring structures are represented by numbers. For instance, the molecule 2-ethyl phenol can be depicted using the following SMILES code: Oc1c(CC)cccc1. Additional symbols are available for special molecules like "/" and "\textbackslash" for cis/trans isomers or "@" for enantiomers.

\begin{figure}[htbp]
    \centering
    \includegraphics[width=0.8\textwidth]{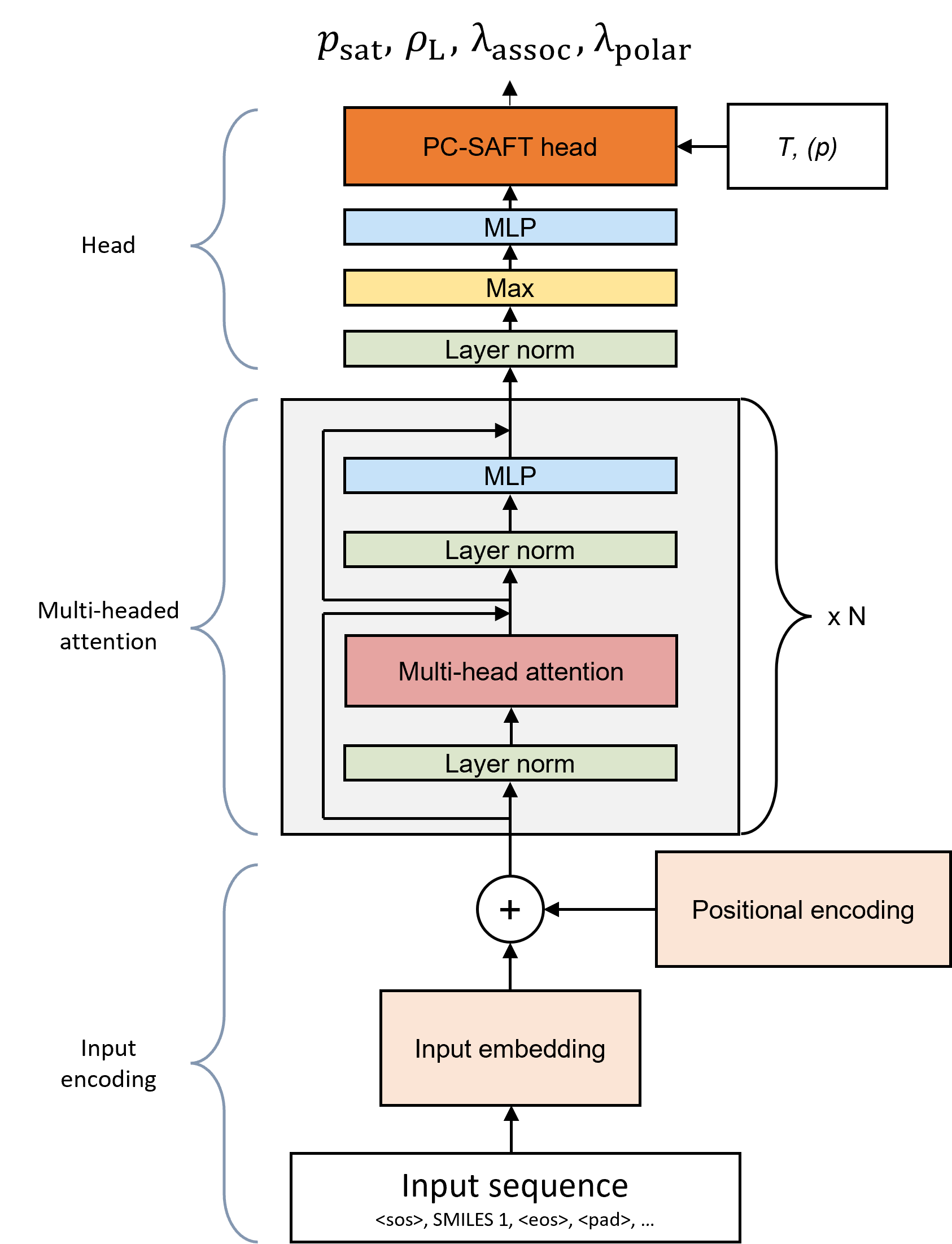}
    \caption{Architecture of \SPTPCSAFT for predicting PC-SAFT parameters using SMILES codes in an end-to-end training. The model takes the SMILES code of a molecule as input. In the input encoding section, information about the individual tokens within the SMILES code and their positions are merged into a single matrix. The multi-head attention section facilitates information exchange between parts of the molecule. In the head section of \SPTPCSAFT, the high-dimensional output from the transformer is first reduced to the number of parameters required by the PC-SAFT head. Subsequently, the output is directed to the PC-SAFT head, which incorporates the PC-SAFT equation of state. The PC-SAFT head receives the temperature $T$ as additional input for the prediction of vapor pressures and the temperature $T$ and the pressure $p$ for the prediction of liquid densities. The outputs of the PC-SAFT head are either vapor pressures and liquid densities as well as association and polarity likelihoods.
}
    \label{fig:network}
\end{figure}

The input of \SPTPCSAFT consists of the SMILES codes representing the molecule of interest with special characters denoting the start of the sequence $<\!\mathrm{SOS}\!>$, and the end of the sequence $<\!\mathrm{EOS}\!>$. The remainder of the input sequence is filled up to a maximum sequence length $n_\mathrm{seq}$ of \num{128} with padding $<\!\mathrm{PAD}\!>$: 
 \begin{equation}
\centering
    <\!\mathrm{SOS}\!>,\mathrm{SMILES},<\!\mathrm{EOS}\!>,<\!\mathrm{PAD}\!>,... \nonumber
    \label{eq:input}
\end{equation}

To render the input string suitable for the machine learning model, the string is tokenized, breaking the sequence into tokens that can each be represented by a unique number. Generally, tokens may comprise multiple characters, but in this work, each token consists of a single character. The tokenization process for SMILES can be compared to assigning first-order groups in group contribution methods. The complete vocabulary containing all tokens can be found in the Supporting Information Section 1.

The input sequence undergoes one-hot encoding, where each token is represented by a learned vector of size $n_\mathrm{emb} = 512$. An input matrix of size $n_\mathrm{emb} \times n_\mathrm{seq}$ is generated by concatenating the vectors representing the tokens of the input sequence. After encoding the input sequence, an additional vector is appended to the right of the input matrix, which holds a linear projection of continuous variables into the embedding space. In the case of the original SPT model~\citep{Winter.2022}, temperature information is encoded in this vector. In \SPTPCSAFT, no continuous variables are supplied here, as temperature and pressure information is only introduced in the final stage (see Figure~\ref{fig:network}), and thus, the continuous variable vector only contains zeros. After adding the continuous variables, the resulting input matrix has a size of $n_\mathrm{emb} \times n_\mathrm{seq}+1$. Subsequently, a learned positional encoding, which contains a learned embedding for each position, of size $n_\mathrm{emb} \times n_\mathrm{seq}+1$ is added to the input matrix. At this stage, the input matrix contains information on all atoms and bonds in the molecule and their positions. However, each token lacks information about its surroundings, as no information has been exchanged between tokens yet. This information sharing between tokens is discussed in the following multi-head attention section.

\subsubsection{Multi-head attention}\label{sec:MHA}

The multi-head attention section sequentially stacks multi-head attention blocks \citep{NIPS2017_3f5ee243}. Within each block, the input undergoes layer normalization before being passed to the multi-head attention mechanism. This mechanism enables information transfer between tokens. Although individual tokens possess only self-information after the input encoding, the multi-head attention mechanism permits tokens to acquire knowledge about their neighbors or other relevant atom or structural tokens within their molecule. Consequently, a transformer block could be viewed as a self-learning n\textsuperscript{th}-order group contribution method, where each token, or the smallest possible group, learns the significance of other tokens and self-assembles higher-order groups based on the molecular structure.

For a more comprehensive and visual explanation, readers are directed to the blog of \citet{Alammar.2018} or the comprehensive description in the Supporting Information of our previous work \citep{winter.2023}.

\iffalse
 
    Mathematically, the output $Z_i$ of a single attention head $i$ is expressed as:
    
    \begin{equation}
    \label{equ:att}
    Z_i = \mathrm{softmax} \left( \frac{Q_i \cdot K_i^\mathrm{T}}{\sqrt{d_k}} \right) V_i
    \end{equation}
    with the query matrix $Q_{i}$, the key matrix $K_{i}$, the value matrix $V_{i}$, and $d_\mathrm{k} = d_\mathrm{emb} / n_\mathrm{head}$, where $n_\mathrm{head}$ represents the number of attention heads. The output $Z_i$ of each head $i$ is concatenated and projected to the size $n_\mathrm{emb} \times n_\mathrm{seq+1}$, which is then passed to a multilayer perceptron (MLP) that concludes the transformer block.
\fi

\subsubsection{The PC-SAFT head}
\label{sec:head}

After the multi-head attention block, the model obtains a high-dimensional representation of the molecule ($n_\mathrm{emb} \times n_\mathrm{seq}$), which needs to be transformed into a set of pure component parameters to be handled within the PC-SAFT equation of state. This dimensionality reduction occurs in the head of the model. We have demonstrated in previous work on the prediction of activity coefficients that it is possible to incorporate physical models like the NRTL equation into the head of our SPT model. However, the PC-SAFT model introduces additional challenges not present in NRTL:

First, the pure component parameters of PC-SAFT have inherent physical meaning, and preserving this physical meaning cannot be guaranteed in a simple regression model. Second, the target properties used for training the model, i.e., vapor pressures and liquid densities, are not direct outputs of PC-SAFT; instead, these target properties must be iteratively converged. While software packages are available that provide robust computations of bulk and phase equilibrium properties with PC-SAFT \citep{Rehner.2023}, it is crucial to ensure that the neural network maintains an intact computational graph to allow the network to obtain a derivative of the target value with respect to all model parameters. An intact computational graph can be ensured when all calculations are conducted within a consistent framework like PyTorch.

%In this work we use the FeO$_\text{s}$ framework to calculate vapor pressures solve the PC-SAFT equation. 

\textbf{Preservation of physical meaning}
   
\begin{figure}[t]
    \centering
    \includegraphics[width=0.8\textwidth]{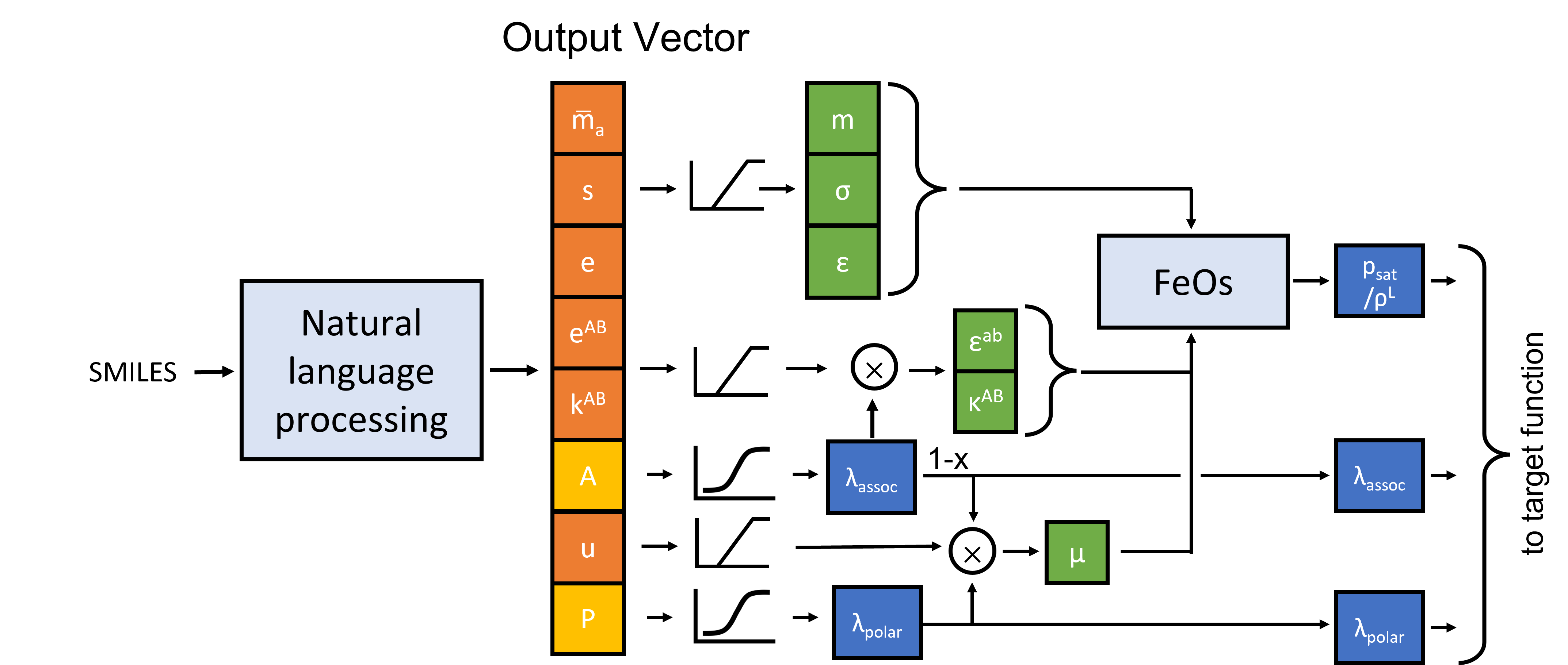}
    \caption{Head section of the model. The natural language processing section of the SPT model returns a vector of length 8. This vector contains six auxiliary pure component parameters of PC-SAFT ($\bar{m}$, $s$, $e$, $e^\mathrm{AB}$, $k^\mathrm{AB}$, and $u$) and the auxiliary association and polarity likelihoods $A$ and $P$. The auxiliary likelihood parameters are passed through a sigmoid function that normalizes them, returning the association and polarity likelihood $\lambda_\mathrm{assoc}$ and $\lambda_\mathrm{polar}$. The associating parameters $\varepsilon^{AB}$ and $\kappa^{AB}$ are calculated by multiplying the auxiliary parameters $e$ and $k$ with $\lambda_\mathrm{assoc}$. The polarity parameter $\mu$ is calculated by multiplying $u$ with $1-\lambda_\mathrm{assoc}$ and $\lambda_\mathrm{polar}$. The resulting pure component parameters are then used in the PC-SAFT equation of state to calculate either vapor pressure or liquid density using the \feos framework \citep{Rehner.2023}. The results of the \feos calculation as well as $\lambda_\mathrm{assoc}$ and $\lambda_\mathrm{polar}$ are passed to the target function.}
    \label{fig:PCSAFT_HEAD}
\end{figure}

The PC-SAFT equation of state is physics-based, and its pure component parameters are related to properties of the underlying molecular model. For example, the pure component parameter $m$ denotes the (potentially non-integer) number of segments on a hypothetical reference fluid, while $\sigma$ and $\varepsilon$ correspond to Lennard-Jones interaction parameters that can be expected to be reasonably transferable between chemically similar molecules. Fortunately, we observe that the pure component parameters $m$, $\sigma$, and $\varepsilon$ naturally converge to subjectively reasonable values. However, this natural convergence is not the case for the pure component parameters that describe polar interactions ($\mu$) and associating interactions ($\varepsilon^{AB}$, $\kappa^{AB}$). These pure component parameters should be 0 for non-polar or non-associating components. This behavior, however, cannot be guaranteed if the parameters are fitted independently by the model purely based on experimental data. Therefore, to ensure the physical meaning of the polar and associating pure component parameters, the \SPTPCSAFT model must learn if a component has associating and polar interactions. To preserve the physical meaning, we predict the polarity and association likelihood in the head of the \SPTPCSAFT model. A graphical description of the PC-SAFT head is given in Fig. \ref{fig:PCSAFT_HEAD}.

After leaving the multi-head attention section, the model has an output of size $n_{emb} \times n_{seq}$. To reduce the dimensionality, a max function is first applied across the sequence dimensions, resulting in a vector of size $n_{emb} \times 1$. Afterward, a linear layer projects this vector to a vector of the auxiliary pure component parameters $\bm{a}$ of size \num{8}, which contains the auxiliary pure component parameters of PC-SAFT ($\bar{m}$, $s$, $e$, $e^\mathrm{AB}$, $k^\mathrm{AB}$, and $u$) and auxiliary association and polarity likelihoods ($A$, $P$). From the auxiliary parameters, the pure component parameters of PC-SAFT $\bm{\phi}$ are calculated using the following equation:

\begin{equation}
\bm{\phi} = \left(1 + \frac{\bm{a}}{10}\right) \cdot \bm{\phi}_{\mathrm{mean}} 
\end{equation}

Here, $\bm{a}$ is the auxiliary parameter, and $\bm{\phi}_{\mathrm{mean}}$ is an externally set hyperparameter determined via a hyperparameter scan. The auxiliary parameters ensure that reasonable values for the pure component parameters of PC-SAFT are reached at the beginning of the training when $\bm{a}$ can be expected to be small values around \num{0}, effectively serving as a staring value for the model. Properly setting the $\bm{\phi}_{\mathrm{mean}}$ parameters ensures quicker convergence. The parameters $m$, $\sigma$, and $\varepsilon$ are now passed directly to the PC-SAFT equation of state, while $\mu$, $\epsilon^\mathrm{AB}$, $\kappa^\mathrm{AB}$ are calculated in the next step.

The calculation of the polar and associating pure component parameters accounts for the polarity and associating likelihoods. To calculate the likelihood, the auxiliary likelihood parameters $A$ and $P$ are passed through a sigmoid function that normalizes them between 0 and 1:
\begin{align}
    \lambda_\mathrm{assoc} &= \frac{1}{1+e^{-A}} \\
    \lambda_\mathrm{polar} &= \frac{1}{1+e^{-P}}
\end{align}

Subsequently, the associating and polar pure component parameters of PC-SAFT are determined by multiplying the likelihood with the auxiliary parameters. For associating molecules, we assume that the association contribution dominates the polar contribution. Thus, the polar pure component parameter is set to 0. Accordingly, the parameters $\mu$, $\varepsilon^\mathrm{AB}$, and $\kappa^\mathrm{AB}$ are calculated as: 

\begin{align}
    \varepsilon^\mathrm{AB} &= e \cdot \lambda_\mathrm{assoc} \\
    \kappa^\mathrm{AB} &= k \cdot \lambda_\mathrm{assoc} \\
    \mu &= u \cdot (1 - \lambda_\mathrm{assoc}) \cdot \lambda_\mathrm{polar}
\end{align}

The parameters $\varepsilon^\mathrm{AB}$, $\kappa^\mathrm{AB}$, and $\mu$ are then passed into the PC-SAFT equation of state to compute either saturation pressures $p^\mathrm{sat}$ or liquid densities $\rho^\mathrm{L}$. The resulting vapor pressures and liquid densities are subsequently passed into the target function along with the associating and polar likelihood $\lambda_\mathrm{assoc}$ and $\lambda_\mathrm{polar}$, respectively. The \SPTPCSAFT setup thus allows the model to learn if components are associating or polar and predict pure component parameters of PC-SAFT with a physical basis.

\textbf{Preservation of the computational graph}

The PC-SAFT equation of state calculates the Helmholtz energy as a function of temperature, mole numbers, and volume. Thermodynamic properties that can be expressed as derivatives of the Helmholtz energy, such as pressure, chemical potential, and heat capacity, are also explicit in terms of temperature, volume, and mole numbers, or, for intensive properties, in temperature $T$ and density $\rho$.

However, the pure component vapor pressure is not directly accessible via a derivative of the Helmholtz energy. Instead, the pure component vapor pressure is implicitly defined as the solution of three nonlinear equations,

\begin{align}
    \mu(T, \rho_\mathrm{V}) &= \mu(T, \rho_{\mathrm{L}}) \\
    p(T, \rho_\mathrm{V}) &= p_\mathrm{sat} \\
    p(T, \rho_\mathrm{L}) &= p_\mathrm{sat}
\end{align}

which need to be solved for the unknown densities $\rho_\mathrm{V}$ and $\rho_\mathrm{L}$, and the vapor pressure $p_\mathrm{sat}$. Fast and robust solvers for this system of equations are implemented in the \feos framework \citep{Rehner.2023} used in this work. However, for the training of the millions of parameters within \SPTPCSAFT, it is mandatory to maintain the full computational graph through the entirety of the neural network, from the output to the input embeddings. If the computational graph is interrupted, derivatives cannot be calculated, rendering learning and thus, training the model impossible. The call to an external program, such as the \feos framework, breaks the computational graph. To address this issue and ensure a fully connected computational graph, we implement the Helmholtz energy calculation of PC-SAFT in PyTorch and conduct the last Newton step of the free energy minimization using the already converged solution from \feos as starting point.

In general, the derivatives of an implicitly defined function $x(\bm{\phi})$ that depends on parameters $\bm{\phi}$ via $f(x, \bm{\phi}) = 0$, can be found by calculating a single step of a Newton iteration starting from an already converged solution $x^\star$ as: 

\begin{equation}
    x(\bm{\phi}) = x^\star - \frac{f(x^\star, \bm{\phi})} {f_x(x^\star, \bm{\phi})}.
\end{equation}

Because $f(x^\star, \bm{\phi})$ is by construction 0, the function value of $x$ does not change. However, due to the explicit dependence on $\bm{\phi}$ automatic differentiation frameworks using both forward mode, in which case $\bm{\phi}$ contains additional dual parts, or backward mode, in which case all operations are recorded on a computational graph, can readily determine the first derivative of $x$ with respect to $\bm{\phi}$.

Applying the concept to the calculation of liquid densities leads to:

\begin{equation}
\rho_\mathrm{L}(T, p, \bm{\phi}) = \rho_\mathrm{L}^* - \frac{p(T, \rho_\mathrm{L}^*, \bm{\phi}) - p}{p_\rho(T, \rho_\mathrm{L}^*, \bm{\phi})}\label{eq:rho_liquid}
\end{equation}

For the vapor pressures, after solving the system of three equations shown above, the last Newton step is:

\begin{equation}
p^\text{sat}(T, \bm{\phi}) = -\frac{a(T, \rho^*_\mathrm{V}, \bm{\phi}) - a(T, \rho^*_\mathrm{L}, \bm{\phi})}{\frac{1}{\rho^*_\mathrm{V}} - \frac{1}{\rho^*_\mathrm{L}}}\label{eq:p_sat}
\end{equation}

with the molar Helmholtz energy $a(T,\rho,\bm{\phi})$. It is particularly convenient that the expression for the vapor pressure only requires an evaluation of the Helmholtz energy in which PC-SAFT and other equations of state are formulated anyway. For liquid densities, however, the pressure and its derivative with respect to density are required. Implementing these derivatives by hand is cumbersome and error-prone. Therefore, we use an additional layer of forward automatic differentiation with second-order dual numbers \citep{Rehner.2021} in which the real and dual parts are PyTorch tensors.

Implementing Eqs. \eqref{eq:rho_liquid} and \eqref{eq:p_sat} into the neural network ensures a fully connected computational graph that can be used by PyTorch to evaluate derivatives of the loss function while still allowing the use of efficient external routines to converge states. While we developed this method to use equations of state, it could also be applied to a wider range of problems where parameters for implicit equations have to be determined using neural networks.

\subsection{Data}
\label{sec:data}

\SPTPCSAFT is trained using vapor pressure and liquid density data obtained from, among others, the Dortmund Data Bank (DDB) \citep{DortmundDatenbank.2022}, the DIPPR database \citep{Thomson.1996} and the ThermoML database \citep{Riccardi.2022} curated by \citet{Esper.2023}.

From this large data collection, all molecules are removed that do not contain at least one carbon atom and most metal complexes except silicon. The remaining data is then split into two sets depending on their data quality: the clean and the remaining dataset. The clean dataset contains molecules that have already been used for the fitting of pure component parameters of PC-SAFT by \citet{Esper.2023} and contains \num{1 103} components, \num{189 504} vapor pressure data points, and \num{282 642} liquid density data points. The pressure data in the clean dataset have undergone a significant effort to eliminate outliers \citep{Esper.2023}. Only data from the clean dataset is used for validation.

The remaining dataset includes the data of the aforementioned databases that is not suitable to directly fit pure component PC-SAFT parameters, as not sufficiently many vapor pressures and liquid densities are available for a given component. However, this data can still be used in \SPTPCSAFT due to the end-to-end training approach. The remaining dataset has a lower data quality than the clean dataset but contains a larger variety of molecules. Several steps were conducted to clean the remaining dataset: First, all data points at a vapor pressure of \qty{1\pm0.01}{\bar} at \qty{298.15\pm1}{\kelvin} are excluded, as these seem to be data points entered erroneously. Then, we removed data points that could not be fitted using PC-SAFT. To remove the data points, we trained eight \SPTPCSAFT models on the clean and remaining data for 15 epochs using a SmoothL1 loss, thus giving less weight to outliers than using an MSE loss. Afterward, we removed all data points from the remaining dataset that have a training loss larger than \num{0.5}. In total, \num{21 456} of \num{233 988} data points were removed from the remaining data. Figure~S1 in the Supporting Information illustrates typical examples of errors identified using our data-cleaning method. Manual review of the removed data points showed that mostly unreasonable-looking data points were removed from the remaining data. Overall, \num{160 186} data points for vapor pressure and \num{52343} data points for liquid densities remain in the data set with \num{12 019} and \num{2 067} molecules each, for vapor pressure and liquid density, respectively.

As our model was employed to clean the remaining data, it is important to note that the remaining dataset is solely used for training the model and not for any form of model validation. For model validation, only the clean dataset is used \citep{Esper.2023}. Thereby, we ensure that our model's performance evaluation is based on reliable and high-quality data and unbiased by our data cleaning steps.

Some of the molecules in the training data are structural isomers such as \textit{cis}-2-butane and \textit{trans}-2-butane. SPT uses isomeric SMILES codes and can thus distinguish between the \textit{cis} and \textit{trans} versions of molecules. However, for some isomeric molecules, our training data also contains data only labeled with the non-isomeric SMILES. In these cases, the data is either one unknown isomer, a mixture of isomers with very similar properties, or mislabeled data of two differently behaving isomers. To avoid ambiguities, we dropped any data related to non-isomeric SMILES codes for components of which isomeric SMILES are present.

\begin{figure}
    \centering
    \includegraphics[width=0.45\textwidth]{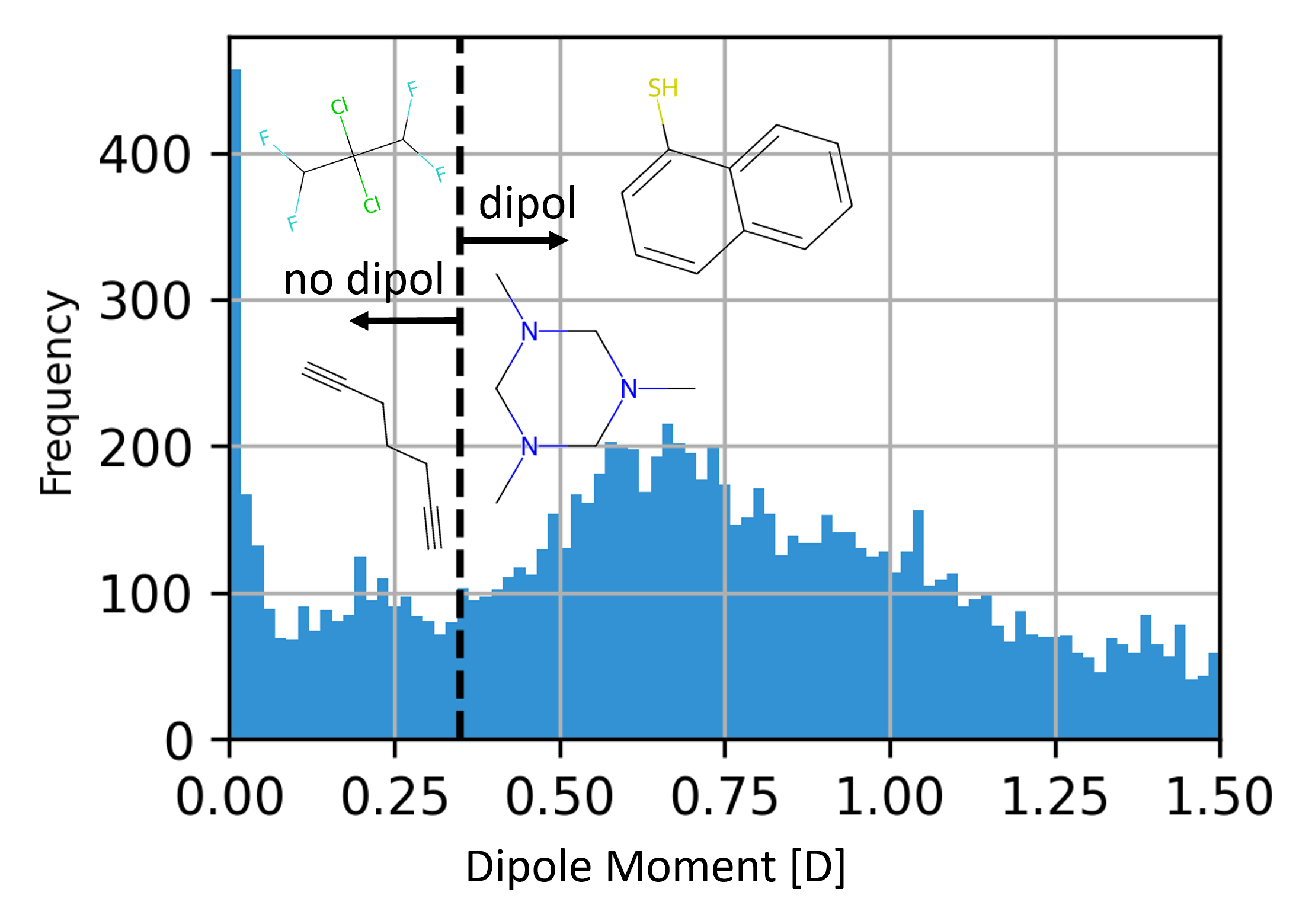}
    \caption{Distribution of dipole moments in the COSMO-Therm database and the threshold of \SI{0.35}{\debye} set to assign polarity. To give a better sense of molecules around the threshold, some molecules with dipole moments close to \SI{0.35}{\debye} are shown. The x-axis represents the range of dipole moments, while the y-axis shows the frequency of molecules in each range.}
    \label{fig:polarity}
\end{figure}

To train the model to recognize if a component is associating or polar, the training data is labeled. To label molecules as associating or polar, we use the following approaches: For associating components, we use RDKit to identify molecules with at least one hydrogen bond donor site and one hydrogen bond acceptor site \citep{GregLandrum.2023}. Components that meet this criterion are labeled as associating. To label molecules as polar, a consistent database of dipole information is needed. Here, we use the COSMO-Therm database 2020, where the dipole moment is available for \num{12 182} molecules in the energy files. If the dipole moment is above \SI{0.35}{\debye}, the molecule is labeled as polar. The limit is set semi-arbitrary by looking at molecules close to the limit and judging if they are polar. Examples of molecules around this polarity threshold are shown in Figure \ref{fig:polarity}. If a component in the training data is unavailable in the COSMO-Therm database, its polarity likelihood is masked in the loss function and thus ignored during training. Polarity information is available for around \SI{95}{\percent} of all molecules in the clean dataset and \SI{50}{\percent} of the molecules in the remaining dataset. \\

\textbf{Validation splits}

In this study, we employ an n-fold cross-validation approach for validating our model using 8 training/validation splits. The data splits are conducted along molecules, ensuring that all data points of a given molecule are either in the training or validation set. This data splitting allows the validation sets to test the model's ability to predict properties of entirely unknown molecules.

However, we impose certain restrictions on the data used for validation. Only components with at least three carbon atoms are included in the validation set, as extrapolation from larger molecules towards very small molecules, such as methane and carbon dioxide, works poorly and the space of small molecules is already well-explored experimentally. Thus, pure component parameters of PC-SAFT are generally available for small molecules \citep{Esper.2023}. Additionally, structural isomers are treated as one component with respect to training/validation splits. Therefore, if the \textit{trans} version of a molecule is in the validation set, the \textit{cis} version is also included in the validation set, and vice versa. The same workflow is applied for enantiomers. 

\subsection{Hyperparameters and training}
\label{sec:hyper}
The base model architecture for \SPTPCSAFT is adopted from our previous SPT-NRTL model \citep{winter.2023} with no further modifications to the architectural hyperparameters such as embedding size, number of layers, and hidden factor. For training \SPTPCSAFT, we use an initial model pretrained on concentration-dependent activity coefficients using a regression head described in \citet{winter.2023}.

To identify good values for $\bm{\phi}_\mathrm{mean}$, we generated a synthetic training dataset with \num{1 494} pure component parameters of PC-SAFT from the work of \citet{Esper.2023} and used these parameters to calculate \num{100} pressure and density values. To validate our model's performance, we reserved \SI{5}{\percent} of the components as a separate validation set. Over this set, a scan was conducted using the parameter values listed in Table \ref{tab:HyperScan}, and the set of parameters leading to the lowest loss on the test set was chosen. 

During the hyperparameter scan, we found that values for $\bm{\phi}_\mathrm{mean}$ that overestimate the critical point help with the convergence. The overestimation ensures that most calculations return valid results in the initial stages of the model training, speeding up the training and avoiding divergence of the model.

\begin{table}[tb]
    \centering
    \caption{Final mean parameter values $\bm{\phi}_\mathrm{mean}$ of the parameter scan. Final values are determined by training a model on a range of parameters and selecting the set of parameters leading to the lowest loss}
    \begin{tabular}{l*{6}{c}}
        \toprule
        Parameter & {$m$} & {$\sigma$ (\si{\angstrom})} & {$\varepsilon/k$ (\si{\kelvin})} & {$\mu$ (\si{\debye})} & {$\kappa^{AB}$} & {$\varepsilon^{AB}/k$ (\si{\kelvin})} \\ 
        \midrule
        $\sigma_\text{mean}$ & 2 & 5 & 300 & 3& 0.005& 1500 \\
        \bottomrule
    \end{tabular}
    \label{tab:HyperScan}
\end{table}

The training was performed on 4 RTX-3090s using a learning rate of \num{1e-4} and 50 epochs. Training takes about \SI{10}{\hour} for 8 training/validation splits running two models per GPU in parallel.

\section{Predictive capabilities of SPT-PC-SAFT}

\label{sec:results}

In our analysis of predictive performance, we utilize the Average Percentage Deviation (APD) as our primary metric. To start, we determine the APD for individual molecules:

\begin{equation}
\text{APD}_i = \frac{1}{M_i} \sum_{j=1}^{M_i} \frac{|y'_{i,j} - y_{i,j}|}{y_{i,j}}
\end{equation}

where \( M_i \) is the number of datapoints for component \( i \), \(y_{i,j} \) is the experimental value and \(y'_{i,j} \) is the predicted value for component $i$ and datapoint $j$. Subsequently, we evaluate either the mean or median of these deviations across the entire dataset. This approach ensures that molecules with numerous data points, such as propane, do not disproportionately influence the prediction discussion. Deviations for vapor pressure $p_\text{sat}$ and liquid density $\rho_\mathrm{L}$ are calculated independently of each other. Unless explicitly stated, we focus on the deviation in the validation set, representing the model's prediction, rather than the deviation in the training set.

\subsection{Prediction of vapor pressures and liquid densities}

\begin{figure}[tb]
    \centering
    \includegraphics[width=0.9\textwidth]{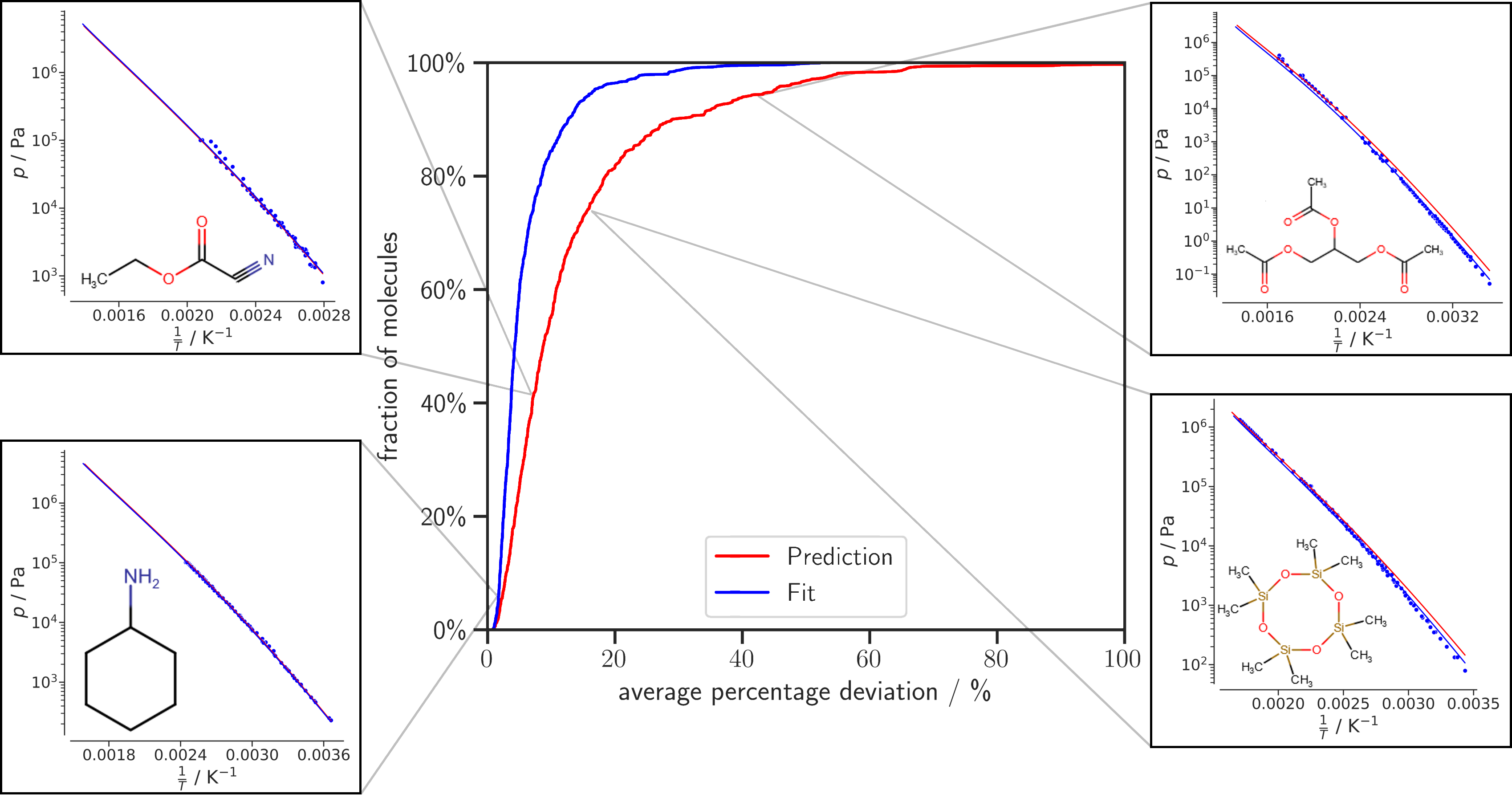}
    \caption{Cumulative deviation curve of vapor pressure prediction of the average percentage deviation for each molecule in our validation set. The fit line represents the average training loss for the same molecules from other splits and serves as a lower bound on the achievable accuracy of our predictive model. To provide a better sense of the APD values, we have included the plot of vapor pressure over $\frac{1}{T}$ for four molecules with APD values of \qty{2}{\percent}, \qty{9}{\percent}, \qty{19}{\percent}, and \qty{45}{\percent}, respectively.}
    \label{fig:resutls_p}
\end{figure}

The \SPTPCSAFT model exhibits a mean APD of \qty{13.5}{\percent} and a median APD of \qty{8.7}{\percent} for predicting vapor pressures in our validation set, consisting of \num{870} components. Figure \ref{fig:resutls_p} presents a cumulative deviation curve of the APD for the validation set and the training set. The training set is comparable to a fitted model and should thus provide an upper bound for the accuracy of PC-SAFT on our training dataset. The results highlight the robustness of \SPTPCSAFT. Only a minor portion of the molecules in the validation set exhibited a notably high APD: \qty{3}{\percent} had an APD exceeding \qty{50}{\percent}, while only \qty{0.4}{\percent} surpassed an APD of \qty{100}{\percent}. This indicates accurate predictions of the vapor pressure for the vast majority of the validation set's molecules.

Figure \ref{fig:resutls_p} illustrates additionally how the APD translates into pressure-temperature ($p$/$T$) plots and demonstrates the diverse set of molecules for which \SPTPCSAFT can account. These examples are cyclohexylamine with an APD of \qty{2}{\percent}, ethyl cyanoacetate with an APD of \qty{9}{\percent}, octamethyl-1,3,5,7,2,4,6,8-tetraoxatetrasilocane with an APD of \qty{19}{\percent}, and triacetin with an APD of \qty{51}{\percent}.

\begin{figure}[tb]
    \centering
    \includegraphics[width=0.45\textwidth]{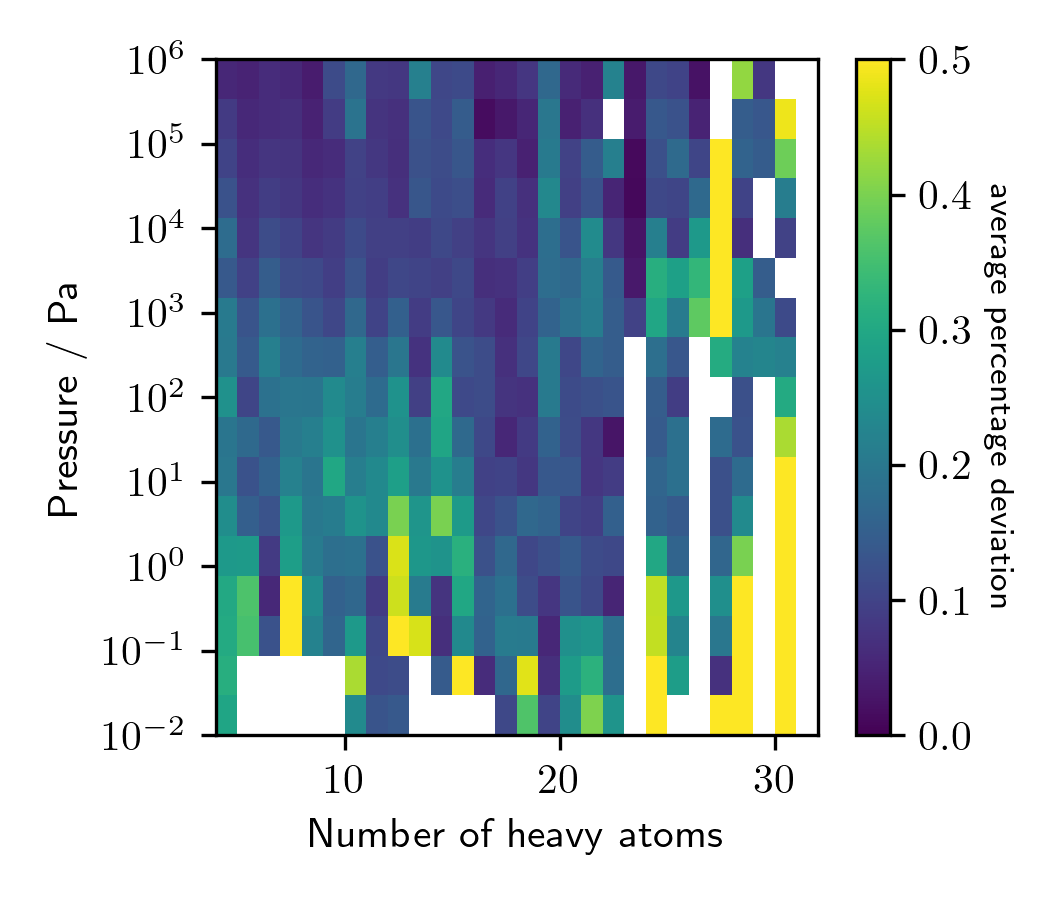}
    \caption{Average percentage deviation in vapor pressure as a function of experimental vapor pressure and the number of heavy atoms in the molecules. Deviations larger than \num{0.5} are truncated at \num{0.5}. }
    \label{fig:heat_p}
\end{figure}

The relationship between APD, molecule size, and vapor pressure range is further illustrated in Figure \ref{fig:heat_p}, which displays the APD in vapor pressure prediction as a function of the number of heavy atoms and pressure. A region of relatively low APD is achieved for molecules containing between 4 and 20 heavy atoms within a vapor pressure range of \qty{1}{\kilo\pascal} to \qty{100}{\mega\pascal}. In contrast, high deviation predominantly occurs at the edges of the data space, particularly for large molecules at low pressures. This behavior might be due to a lower density of data and higher uncertainty when measuring low-pressure systems.

\begin{figure}[bt!]
    \centering
    \includegraphics[width=\textwidth]{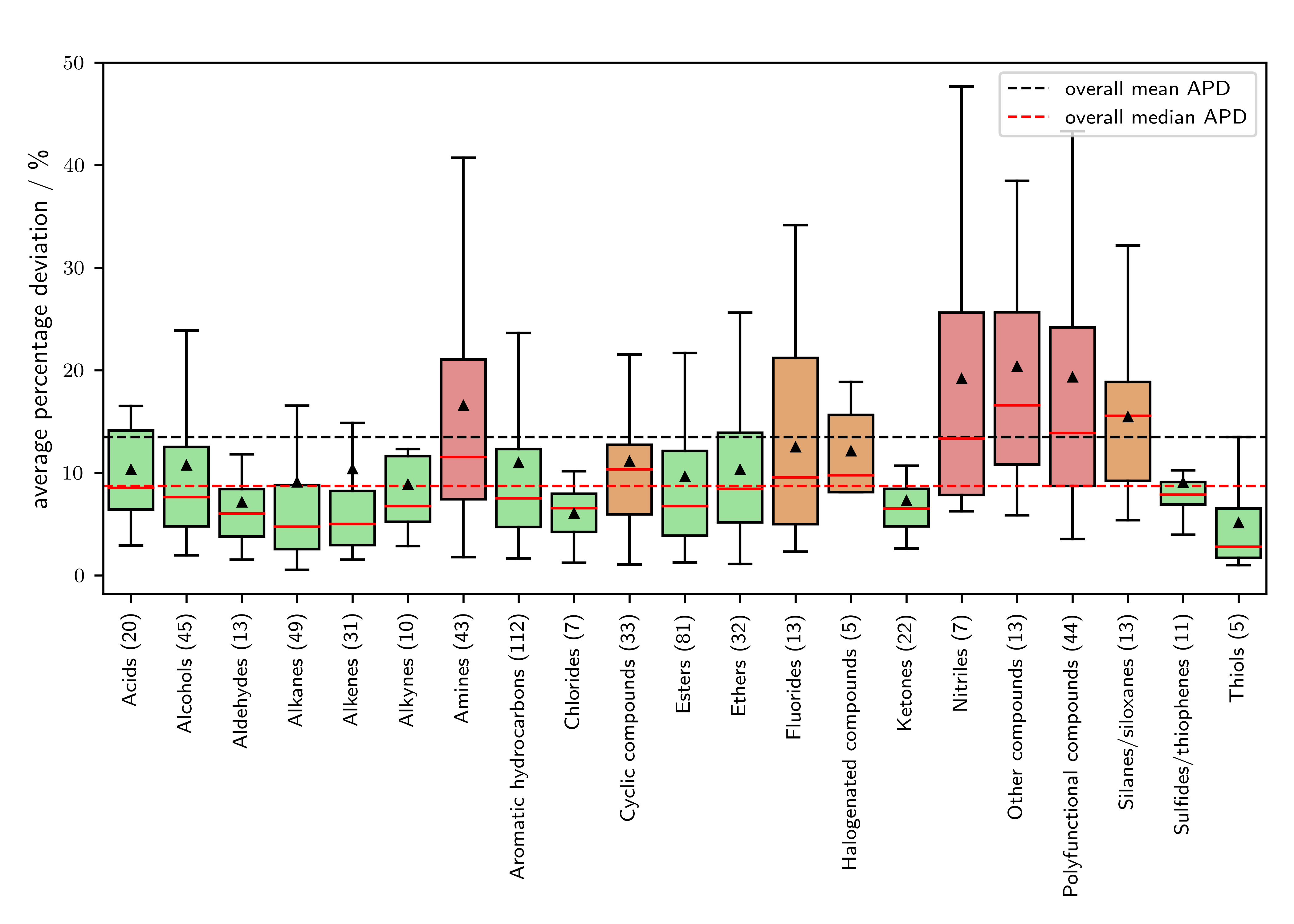}
    \caption{Average percentage deviation in vapor pressure as a function of the molecular family. Molecular families are assigned according to the DIPPR database \citep{Thomson.1996}. Of the \num{870} components in the validation set, \num{609} components could be assigned a molecular family. Green boxes show families with a median APD of \qty{2.5}{\percent} below the overall mean APD of \qty{13.5}{\percent}, red boxes show families with an APD of \qty{2.5}{\percent} above the overall  mean APD.}
    \label{fig:fammily}
\end{figure}

In Figure \ref{fig:fammily}, the relationship between APD (Average Percentage Deviation) and molecular families is explored. The classification of the molecular families is based on the DIPPR database \citep{Thomson.1996}, which contains families for \num{609} out of the \num{870} components in the validation set. A noticeable correlation is obtained between the expected prediction error and the molecular families. Notably, molecular families composed solely of oxygen and carbon exhibit above-average prediction accuracy. In contrast, fluorinated, halogenated (bromide and iodine), and particularly nitrogen-containing compounds present challenges in prediction. A comprehensive list of the validation set, categorized by molecular group, can be found in the Supplementary Information (SI). Overall \SPTPCSAFT, performs well for the majority of molecular families.

\begin{figure}[tb]

    \centering    
    \captionsetup[subfigure]{oneside,margin={1cm,0cm}}
    \subfloat[alkanes]{
    \includegraphics[width=0.45\textwidth]{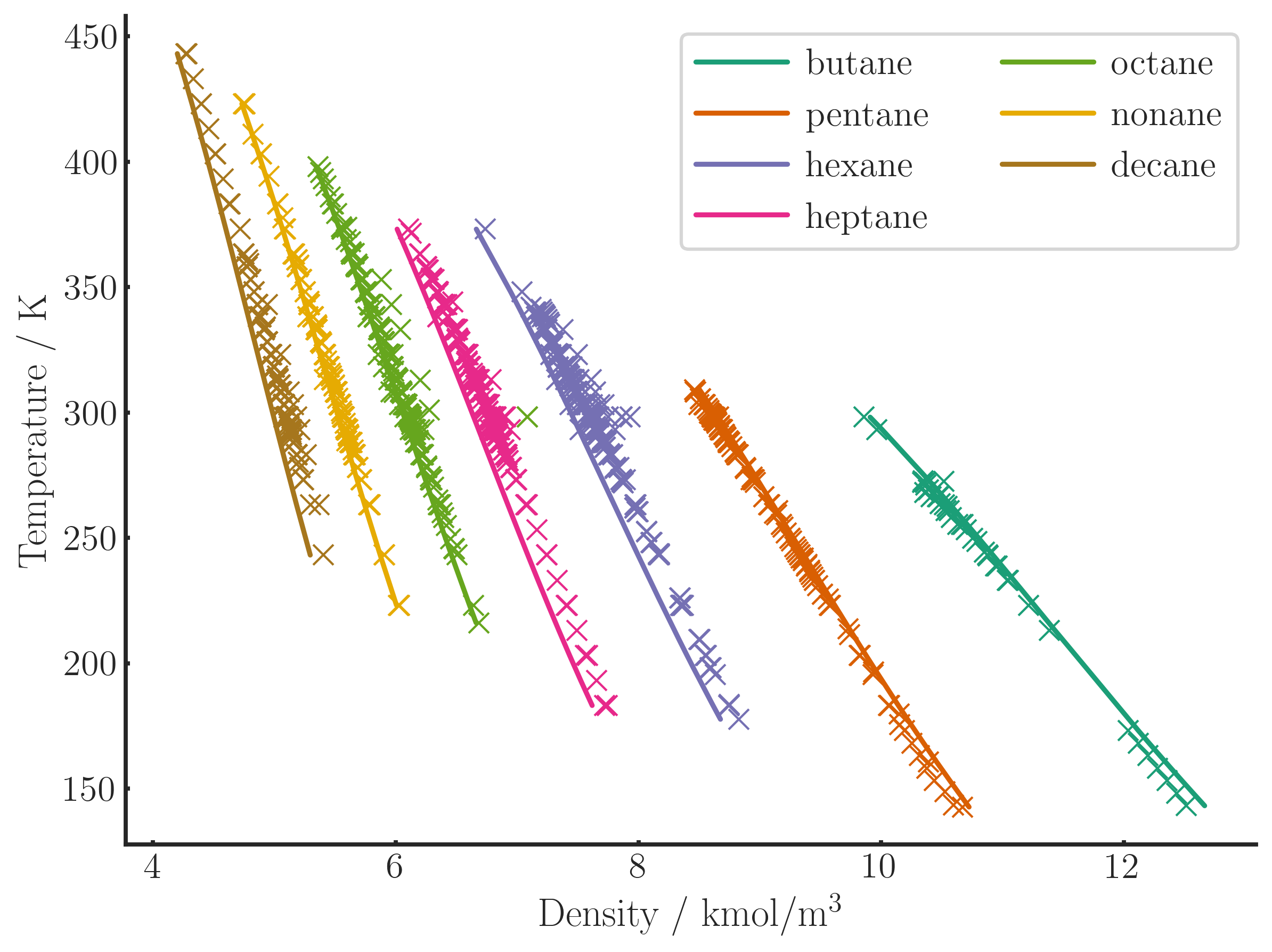}%
    \label{fig:rho_alkane}
    }    
    \centering
    \captionsetup[subfigure]{oneside,margin={1cm,0cm}}
    \subfloat[alcohols]{
    \includegraphics[width=0.45\textwidth]{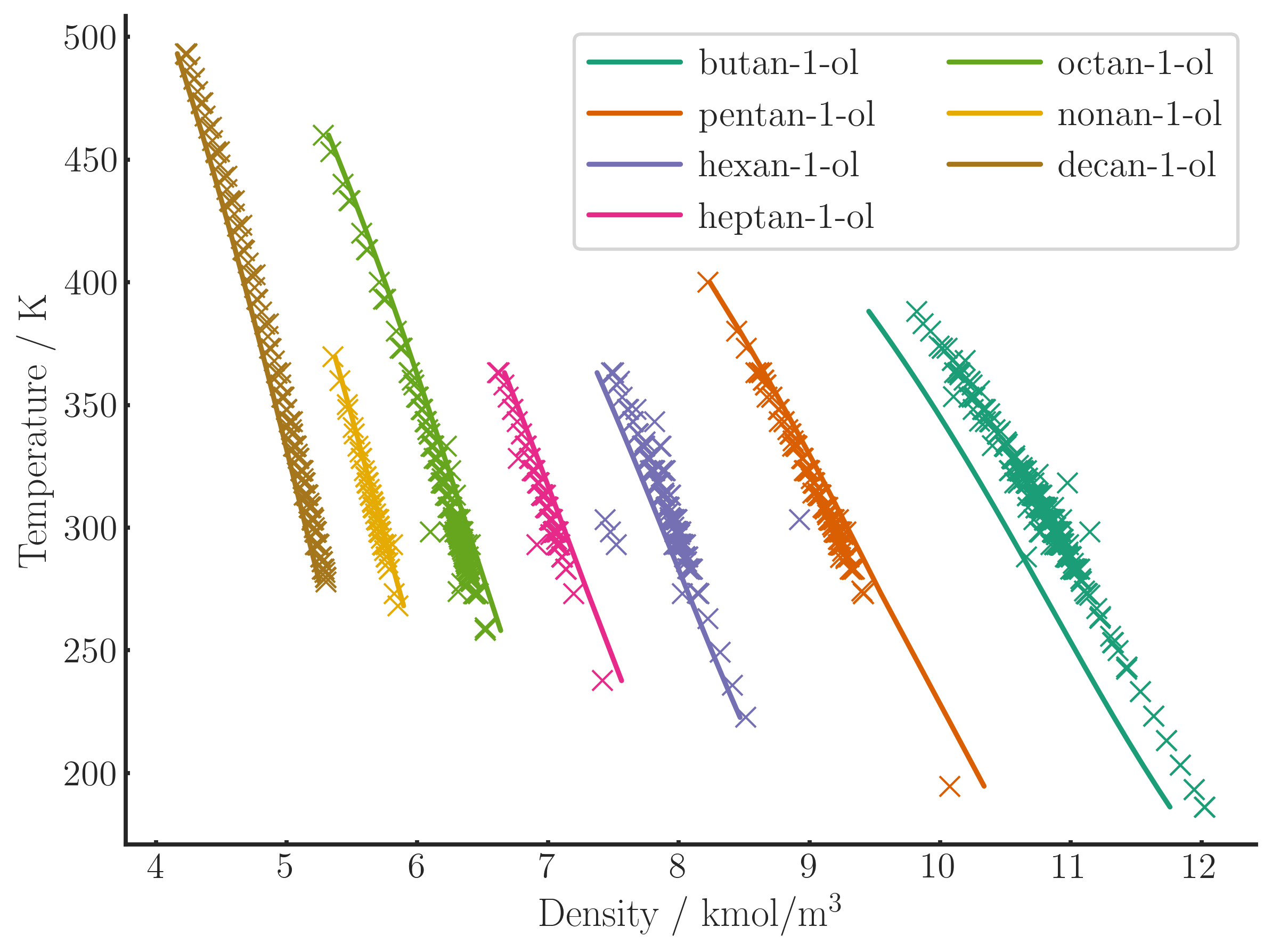}%
    \label{fig:rho_alcohol}
    }

    \caption{Prediction of molar density of C4 to C10 alkanes and alcohols at \qty{1}{\bar} over a range of temperatures using \SPTPCSAFT (lines). Experimental data (crosses) are taken from the DDB. }
    \label{fig:rho}
\end{figure}

The APD in liquid density is generally lower than the deviation in vapor pressure. For densities, our \SPTPCSAFT model achieves a mean APD of \qty{3.1}{\percent}. Predicted liquid densities at \qty{1}{\bar} are shown for a range of alkanes and alcohols in Figure~\ref{fig:rho}, generally demonstrating a good agreement with the measured data.

\subsection{Physicality of predicted pure component parameters of PC-SAFT}

One major advantage of the PC-SAFT model is the physical basis of its parameters. Thus, any predictive model should retain this physicality. We preserve the physical meaning of the predicted pure component parameters by introducing the polarity and association likelihood (see Section~\ref{sec:head}). Table \ref{tab:parameters} provides an overview of selected pure component parameters of PC-SAFT predicted by \SPTPCSAFT. The pure component parameters $m$, $\sigma$, and $\varepsilon$ are predicted within anticipated ranges. The chain length parameter $m$ increases along the homologous series, while the segment diameter $\sigma$ and interaction energy $\varepsilon$ are similar for molecules in the same chemical family. The association is accurately identified for alcohols, and polarity is properly assigned to ethers. On the one hand, 1-ethoxypentane gets assigned a dipole moment of \SI{2.5}{\debye} with a polarity likelihood of nearly 1. On the other hand, 1,2-diethoxymethane exhibits no dipole moment due to its higher symmetry, as correctly recognized by \SPTPCSAFT. Consequently, the predicted parameters seem physically plausible. 

The Supplementary Information (SI) presents the receiver operating characteristic (ROC) curves of the association and polarity likelihood parameters, illustrating the trade-off between true positives and false positives. \SPTPCSAFT achieves a \qty{100}{\percent} true positive rate for associating molecules and approximately a \qty{90}{\percent} true positive rate for polarity. Given that we use classification in the normally continuous spectrum for polarity, a \qty{100}{\percent} true positive rate is not expected. Therefore, our model architecture enables \SPTPCSAFT to accurately learn when molecules exhibit associating or polar interactions and assign appropriate pure component parameters.

\begin{table}[tb]
    \centering
    \caption{Examples of pure component PC-SAFT parameters predicted by \SPTPCSAFT.}
    \begin{tabular}{llSSSSSS}
        \toprule
        Name & SMILES  & {$m$} & {$\sigma$ (\si{\angstrom})} & {$\varepsilon/k$ (\si{\kelvin})} & {$\mu$ (\si{\debye})} & {$\kappa^{AB}$} & {$\varepsilon^{AB}/k$ (\si{\kelvin})} \\ 
        \midrule
        butane &CCCC &2.3 &3.7 &224 & & & \\
        hexane &CCCCCC &2.9 &3.9 &244 & & & \\
        octane &CCCCCCCC &3.6 &3.9 &248 & & & \\
        1-butanol &CCCCO &3.2 &3.5 &247 & &0.006 &2409 \\
        1-hexanol &CCCCCCO &3.7 &3.6 &258 & &0.005 &2498 \\
        1-ethoxypentane &CCCCCOCC &3.9 &3.7 &236 &2.5 & & \\
        1,2-diethoxymethane &CCOCOCC &3.6 &3.5 &231 &  & & \\
        \bottomrule
    \end{tabular}
    \label{tab:parameters}
\end{table}

\subsection{Comparison to homosegmented GC method and recent ML models}

To assess the predictive capabilities of our method, we compare it to the homo-segmented group contribution method proposed by \citet{Sauer.2014}, in the following called GC-Sauer. The group contribution method by \citet{Sauer.2014} calculates the PC-SAFT parameters from the contributions of individual functional groups. We define two sets of molecules that differ in the breadth of the molecular space: The \textit{interpolation set} contains molecules that belong to the chemical families that \citet{Sauer.2014} used to parameterize the GC method (branched alkanes, alkenes, 1-alkynes, alkylbenzenes, alkylcyclohexanes, alkylcyclopentanes, ethers, aldehydes, formates, esters, ketones, 1-alcohols, and 1-amines) but only containing a maximum of one functional group as in \citet{Sauer.2014}. The \textit{interpolation set} likely contains many of the molecules on which the GC method was originally fitted. Thus, the GC-Sauer method enjoys a maximum advantage in the comparison. The \textit{extrapolation set} contains molecules outside of these chemical families that can still be fragmented into the groups defined by \citet{Sauer.2014} but that do not contain more than one polar or associating group to not extrapolate from the GC-Sauer method to far. The \textit{extrapolation set} contains important molecules like cyclohexylamine or phenyl acetate that are difficult to describe accurately for GC methods. In total, the \textit{interpolation set} contains \num{256} molecules and the \textit{extrapolation set} contains \num{67} molecules.

\begin{figure}[bt]

    \centering    
    \captionsetup[subfigure]{oneside,margin={1cm,0cm}}
    \subfloat[Vapor pressure $p_\mathrm{vap}$]{
    \includegraphics[width=0.4\textwidth]{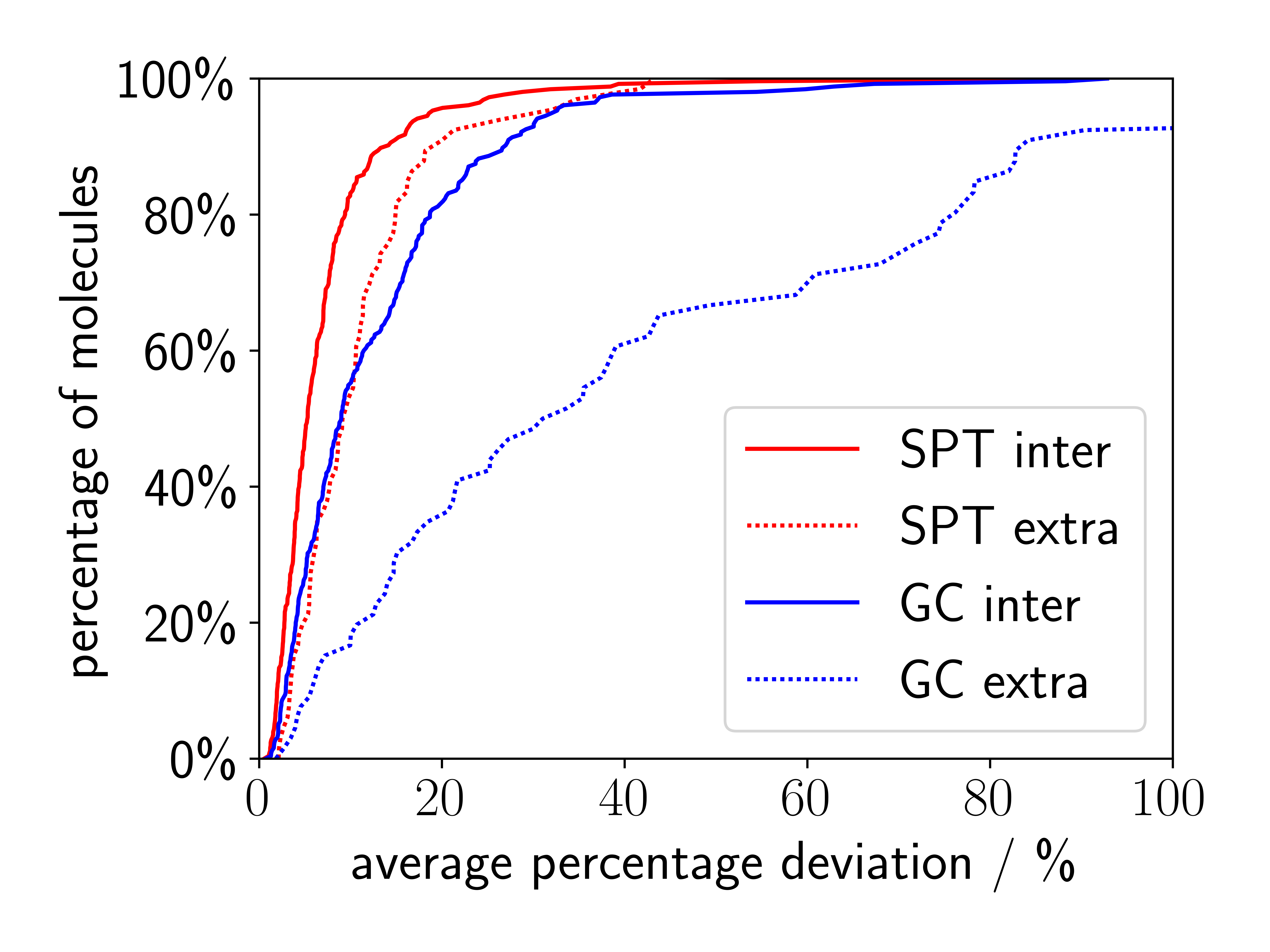}%
    \label{fig:GC_SAUER_p}
    }    
    \centering
    \captionsetup[subfigure]{oneside,margin={1cm,0cm}}
    \subfloat[Liquid density $\rho_\mathrm{L}$]{
    \includegraphics[width=0.4\textwidth]{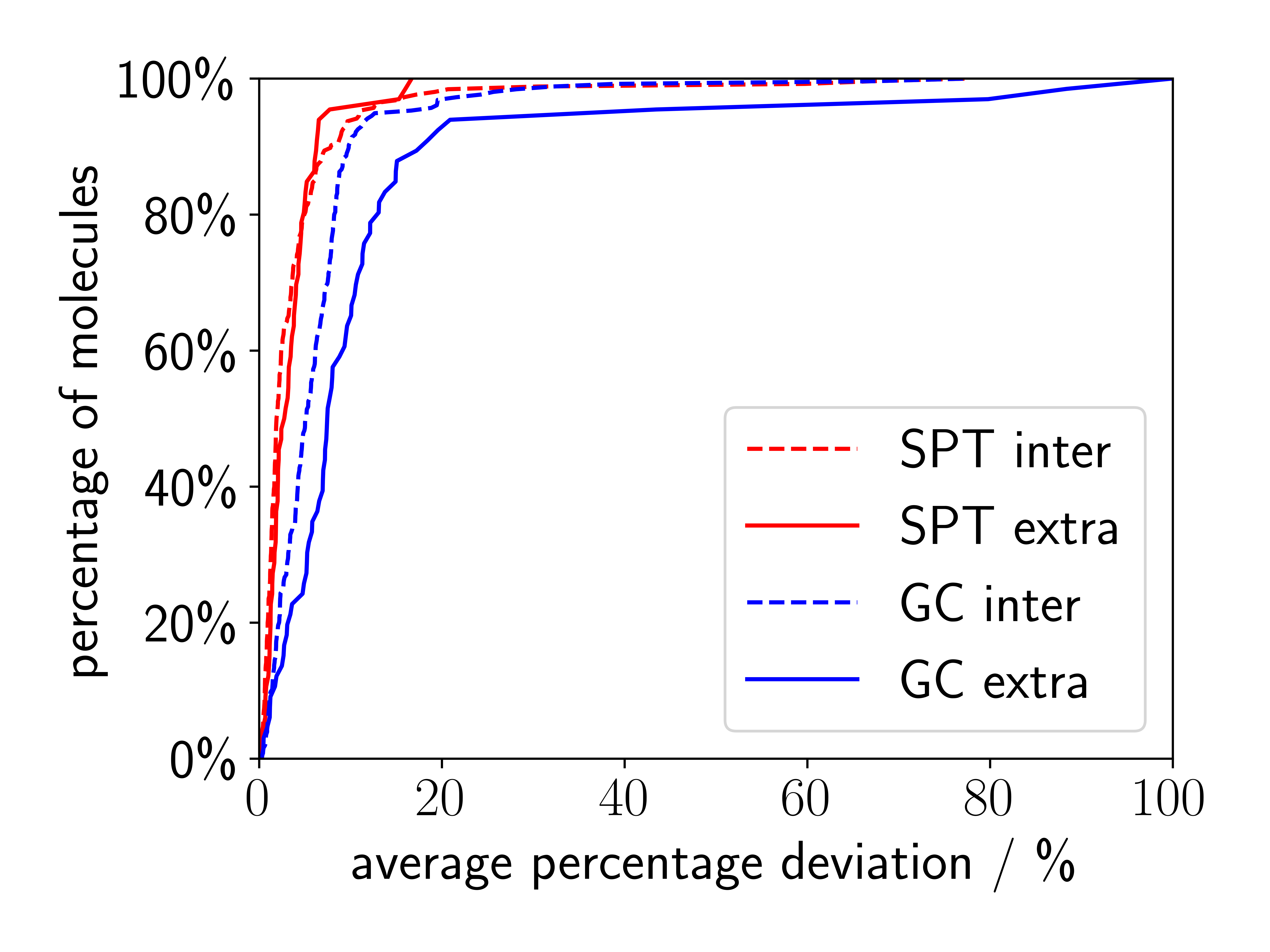}%
    \label{fig:GC_SAUER_v}
    }

    \caption{Cumulative deviation plot of the average percentage deviations of the molecules in the extrapolation and \textit{interpolation set}s for predictions of (a) vapor pressures $p^\mathrm{vap}$ and (b) liquid densities $\rho^\mathrm{L}$. The predictive performance of both models is lower on the extrapolation dataset, where SPT outperforms GC-Sauer significantly.}
    \label{fig:GC_SAUER}
\end{figure}

The comparison between \SPTPCSAFT and GC-Sauer on the two sets of molecules indicates a substantial difference between the performance of the GC-Sauer and \SPTPCSAFT methods when extrapolating beyond the \textit{interpolation set} (Figure \ref{fig:GC_SAUER}): While the GC method performs decently within the \textit{interpolation set}, with a mean APD of \qty{12.8}{\percent} compared to \qty{7.3}{\percent} of \SPTPCSAFT for the vapor pressure, it falls short when extrapolating to more complex molecules, resulting in a much larger mean APD of \qty{48.0}{\percent} compared to \qty{11.1}{\percent} for \SPTPCSAFT. Similar performance benefits are observed for \SPTPCSAFT in predicting liquid densities. Here, for the \textit{interpolation set}, \SPTPCSAFT has an mean APD of \qty{4.0}{\percent} compared to \qty{6.4}{\percent} of GC-Sauer and, for the \textit{extrapolation set}, \qty{3.5}{\percent} compared to \qty{11.9}{\percent} of GC-Sauer.

Compared to the recently published methods by \citet{Felton.2023} and \citet{Habicht.2023}, \SPTPCSAFT compares favorably. However, since there is no consistent validation set used across the studies, there is some uncertainty in this discussion. The reported average relative percentage errors in vapor pressures by \citet{Felton.2023} are \qty{39}{\percent}  based on a similar dataset as our clean dataset, compared to \SPTPCSAFT mean APD of \qty{13.5}{\percent}. \citet{Habicht.2023} report average relative percentage deviations below \qty{20}{\percent} for many molecular families, however, limited to non-polar, non-associating molecules for which \SPTPCSAFT has a mean deviation of \qty{10}{\percent}. Overall, the better performance of \SPTPCSAFT might lie in the direct training on experimental data and not on previously fitted PC-SAFT parameters. Thus, \SPTPCSAFT is able to use a larger amount of data points and avoids error accumulation via the additional regression step.

Our results demonstrate that the much simpler GC method of \citet{Sauer.2014} performs reasonably well for molecules similar or equal to those to which it was parameterized, but extrapolating capabilities are limited for more complex molecules. To cover a more comprehensive molecular space without manually defining an extensive set of (potentially higher order) groups, an approach that captures the complexities of molecules, like \SPTPCSAFT, is required. Moreover, even compared to more complex and recent machine learning approaches \SPTPCSAFT compares favorably. 

\subsection{Differentiation of stereoisomers}

\begin{figure}[tb]
    \centering   
    \captionsetup[subfigure]{oneside,margin={1cm,0cm}}
    \subfloat[1,1,1,4,4,4-hexafluorobutene]{
    \includegraphics[width=0.4\textwidth]{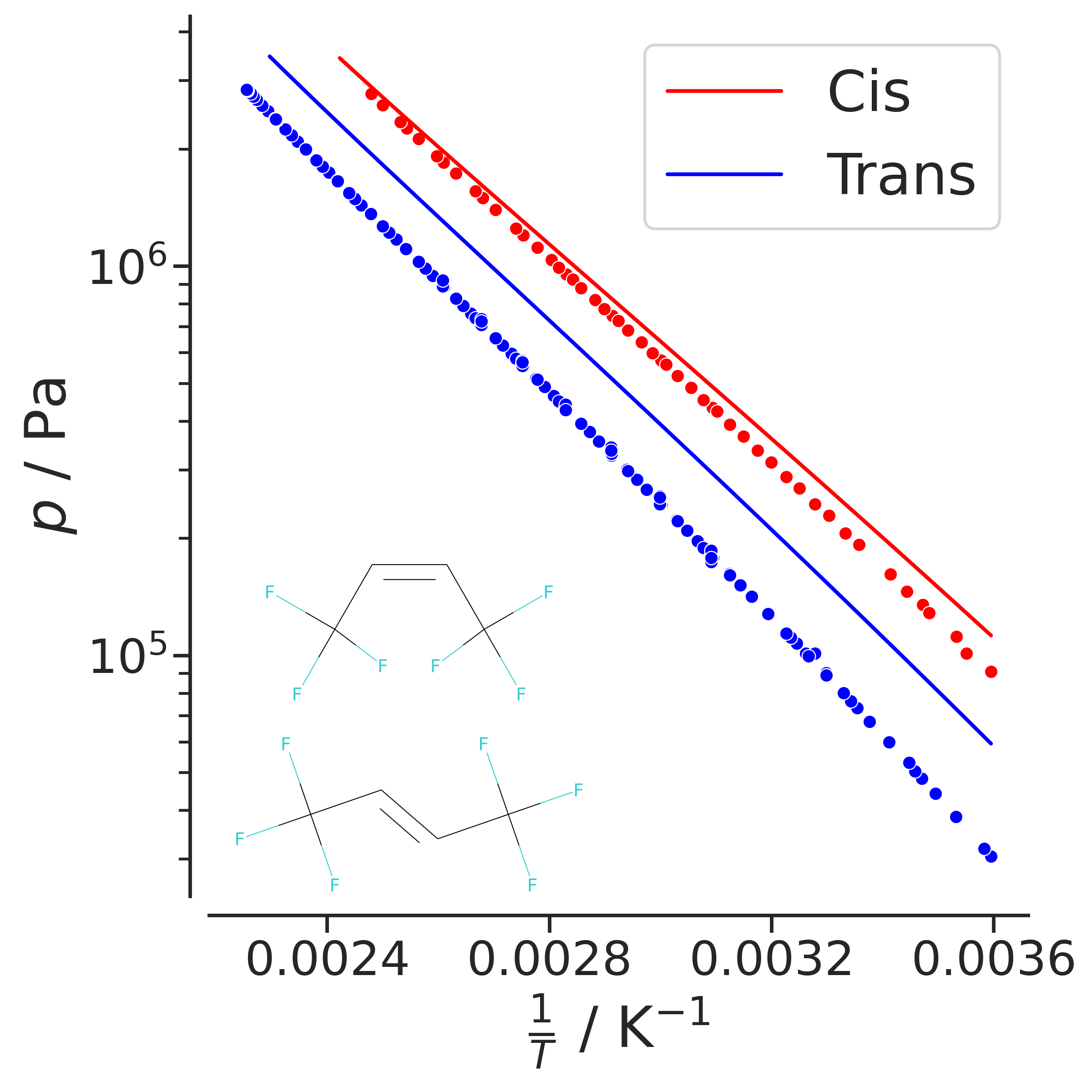}%
    \label{fig:isomers1}
    }    
    \centering
    \captionsetup[subfigure]{oneside,margin={1cm,0cm}}
    \subfloat[stilbene]{
    \includegraphics[width=0.4\textwidth]{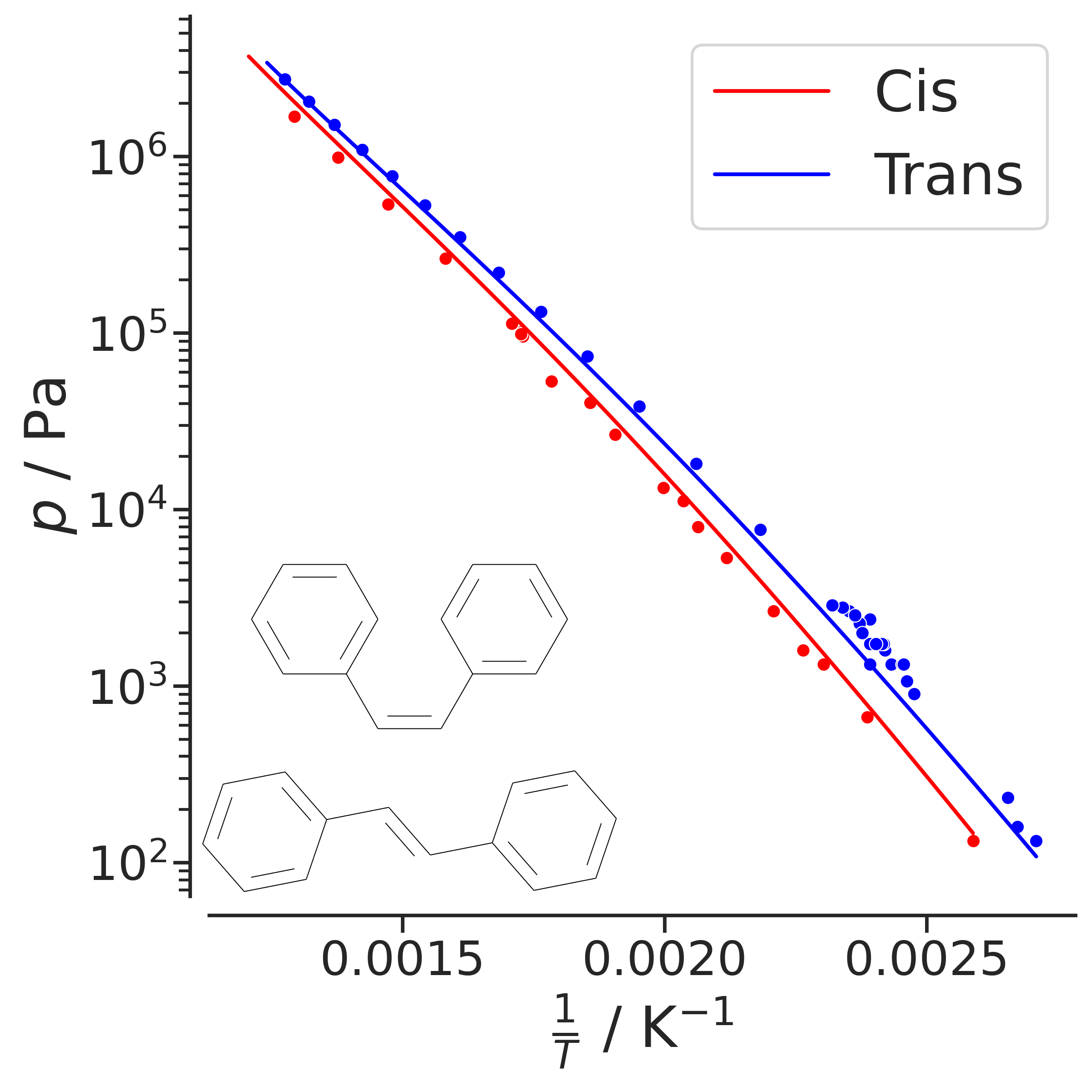}%
    \label{fig:isomers2}
    }
    \centering
    \captionsetup[subfigure]{oneside,margin={1cm,0cm}}
    \subfloat[2-hexene]{
    \includegraphics[width=0.4\textwidth]{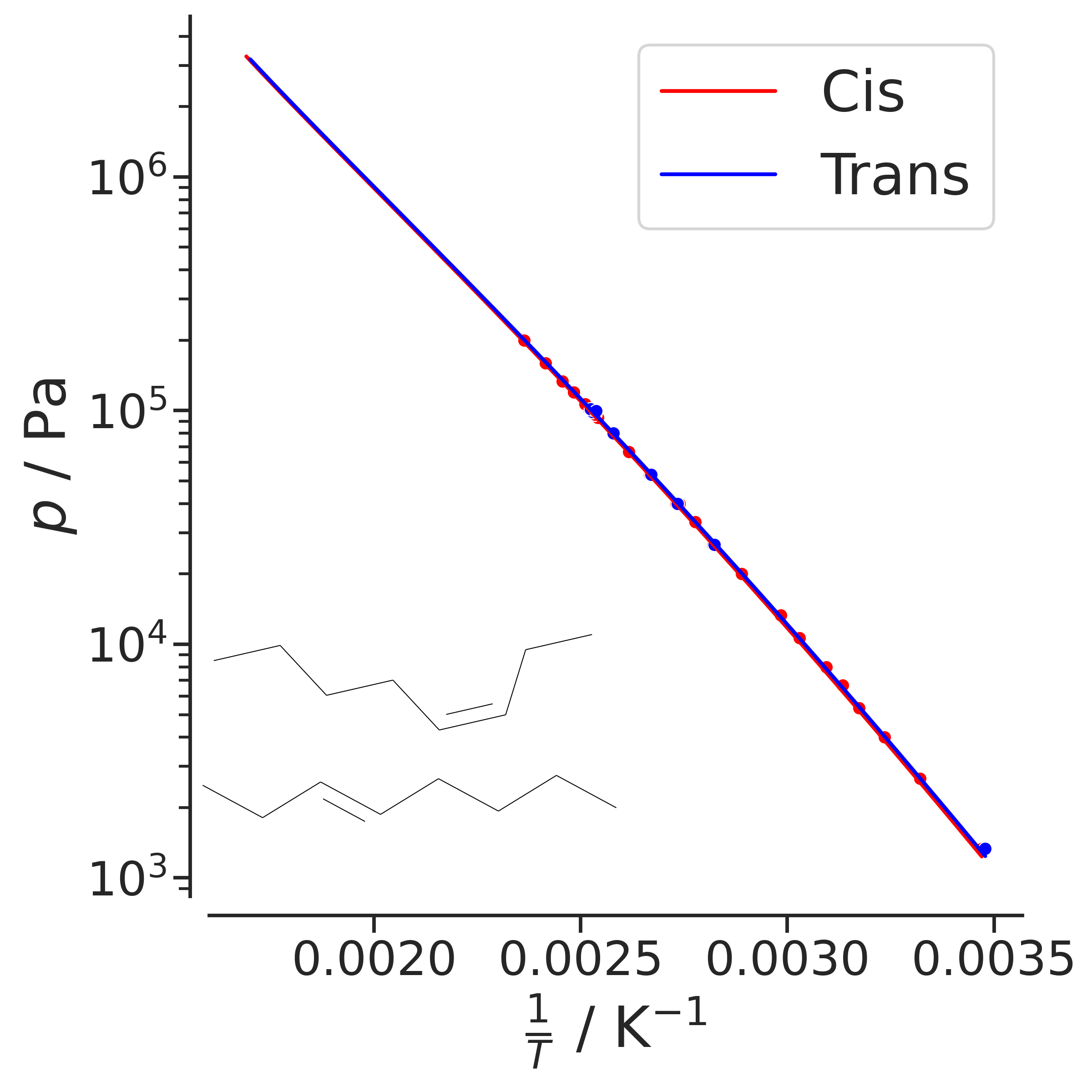}%
    \label{fig:isomers3}
    }
    \centering
    \captionsetup[subfigure]{oneside,margin={1cm,0cm}}
    \subfloat[2-hexenedinitril]{
    \includegraphics[width=0.4\textwidth]{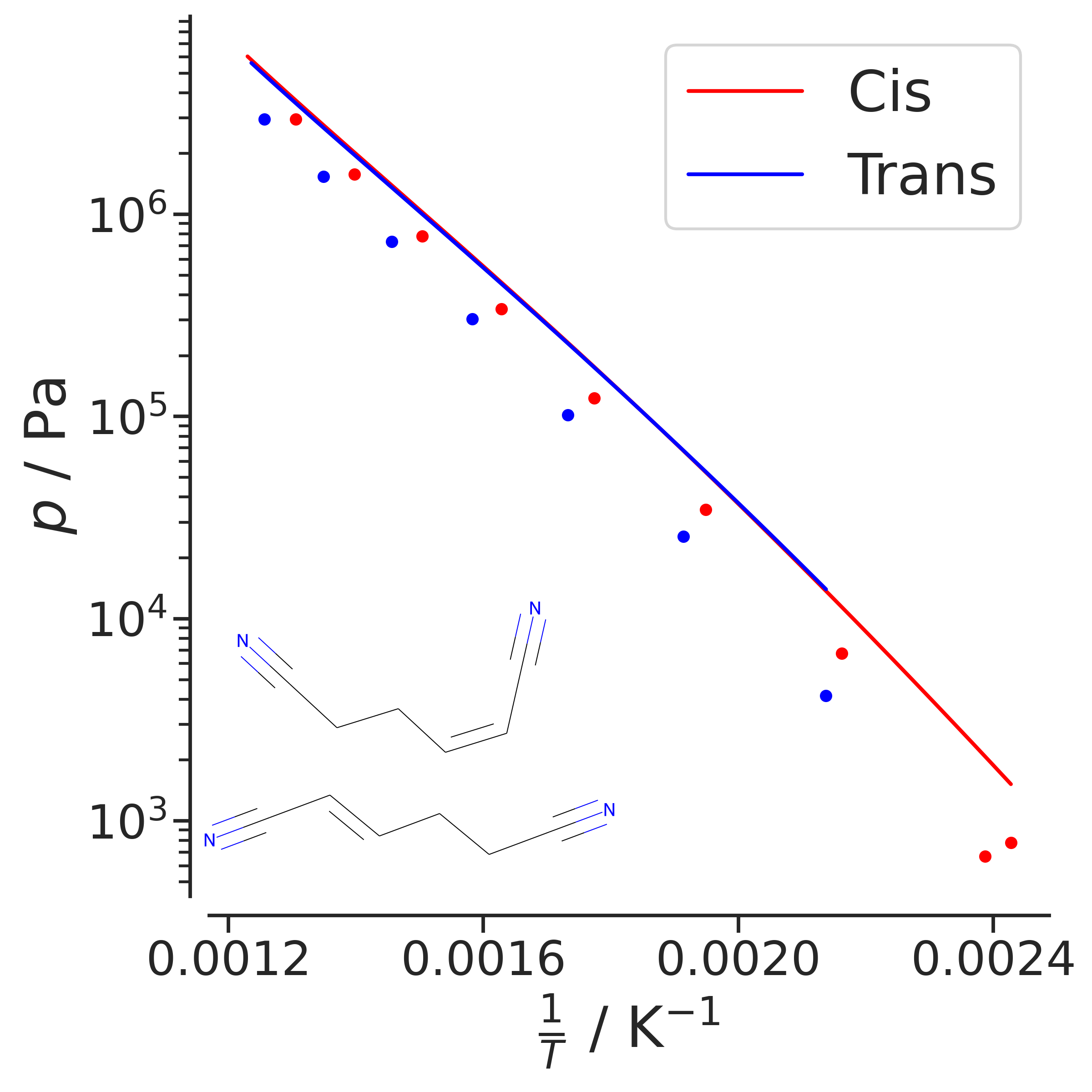}%
    \label{fig:isomers4}
    }
    \caption{Pressure-temperature plots of the isomer pairs 1,1,1,4,4,4-hexafluorobutene, stilbene, 2-hexene and 2-hexenedinitril}
    \label{fig:isomers}
\end{figure}

Stereoisomers are molecules that have the same molecular formula and constitution but different structural arrangements due to differently arranged bonds. Although these subtle structural differences might appear insignificant, they can impact the properties of isomers substantially in some cases. GC methods often struggle to capture these differences in stereoisomers as they require large higher-order groups to differentiate between them. However, \SPTPCSAFT utilizes isomeric SMILES as input, enabling the model to distinguish between stereoisomers. Unfortunately, our validation data contains only 35 pairs of stereoisomers, the majority of which exhibit no significant difference in vapor pressure. Therefore, we assess the prediction of stereoisomers based on individual examples and a comprehensive statistical analysis has to be performed as soon as more data on stereoisomers is available. 

For four example isomere pairs, i.e., the \textit{cis} and \textit{trans} isomers of 1,1,1,4,4,4-hexafluorobutene, stilbene, 2-hexene, and 2-hexenedinitril, the predicted vapor pressure is shown in Figure \ref{fig:isomers}. Due to the different polarity, the isomers of 1,1,1,4,4,4-hexafluorobutene and stilbene have measurably different vapor pressures. \SPTPCSAFT is able to predict the trend in vapor pressures, which is remarkable considering that the majority of isomers in the training data is similar to 2-hexene which shows no significant difference between the two isomers. However, 2-hexenedinitrile presents a challenge for the model, as it fails to distinguish between isomers even though there is a difference in vapor pressure between the \textit{cis} and \textit{trans} versions. When and why \SPTPCSAFT fails in distinguishing specific isomers should be subject to further research. We observed some instances within our training data of likely mislabeling between isomers, which may impede the model's performance. Overall, the results concerning stereoisomer differentiation are encouraging, but more and better data on stereoisomers is required to unlock the full capability of the model.

\subsection{Publication of predicted pure component parameters}

While the current \SPTPCSAFT model is efficient and straightforward to set up, executing machine learning models can still present a barrier to entry when only single components are of interest. To enhance the accessibility of our model, we have predicted pure component parameters of PC-SAFT for millions of components, as we have previously with a set of 100 million NRTL parameters \citep{winter.2023}. Predicted pure component parameters of PC-SAFT are available for all \num{13645} molecules contained in our training set.

By making these pre-computed pure component parameters available, we aim to facilitate broader adoption and utilization of the PC-SAFT equation of state across various applications and allow for exploring vast molecular spaces.

\section{Conclusion}
\label{sec:conclusion}

In this study, we introduce the machine-learning model \SPTPCSAFT, which can predict thermodynamic equilibrium properties using the PC-SAFT equation of state and the corresponding pure component parameters of PC-SAFT from the SMILES code of a molecule. \SPTPCSAFT is a modification of the SMILES-to-Properties-Transformer (SPT) \citep{Winter.2022} and overcomes challenges posed by the complexity of the PC-SAFT equation of state while preserving the physical meaning of its parameters. 

Our model demonstrates excellent predictive performance on a validation set of \num{870} components, achieving a mean APD of \qty{13.5}{\percent} for vapor pressures and \qty{3}{\percent} for liquid densities. Remarkably, \qty{99.6}{\percent} of the predictions fall within a factor of 2, indicating a minimal presence of outliers.

Compared to the homo-segmented group contribution method of PC-SAFT by \citet{Sauer.2014}, our \SPTPCSAFT model provides significantly higher quality predictions for both vapor pressures and liquid densities and compares favorably to more recent ML models. In particular, for more complex molecules, the prediction accuracy of \SPTPCSAFT is four times higher than the group contribution method. Moreover, our model can differentiate between stereoisomers, highlighting its potential for improved accuracy in predicting the properties of subtle molecular effects. We believe that \SPTPCSAFT offers a versatile and robust approach for predicting equilibrium thermodynamic properties and the corresponding pure component parameters of PC-SAFT, allowing for applications in thermodynamics, process engineering, and material science.

To make our model more accessible to researchers and industry professionals, we have precomputed pure component parameters of PC-SAFT for a large number of components. 

The \SPTPCSAFT model presents a significant advancement in the prediction of equilibrium properties and corresponding pure component parameters of PC-SAFT. By leveraging machine learning techniques, our model offers improved accuracy in predicting the properties of various molecules while being capable of handling complex molecular structures and subtle differences in isomers. The availability of precomputed pure component parameters of PC-SAFT will further facilitate the adoption of our model and enable its use in a broad range of research and industry applications.

\section*{Acknowledgments}

B.W. and A.B. acknowledge funding by NCCR Catalysis, a National Centre of Competence in Research funded by the Swiss National Science Foundation, grant number 180544.

P.R. acknowledges funding by the Deutsche Forschungsgemeinschaft (DFG, German Research Foundation), grant number 497566159.

%Bibliography
\bibliographystyle{abbrvnat}  
\bibliography{references}

\end{document}